\documentstyle[12pt]{article}
\newcommand{\be}{\begin{eqnarray}}
\newcommand{\ee}{\end{eqnarray}}
\newcommand{\ba}{\begin{array}}
\newcommand{\ea}{\end{array}}
\newcommand{\half}{{\textstyle{\frac{1}{2}}}}
\newcommand{\textfrac}[2]{{\textstyle{\frac{#1}{#2}}}}
\newcommand{\fourint}[1]{\int\!\frac{d^4 #1}{(2\pi)^4}}
\newcommand{\Fdual}{\widetilde{F}}
\newcommand{\cF}{{\cal F}}
\newcommand{\cR}{{\cal R}}
\newcommand{\cO}{{\cal O}}
\newcommand{\cM}{{\cal M}}
\newcommand{\cD}{{\cal D}}
\newcommand{\tr}{{\rm tr}\,}
\newcommand{\Tr}{{\rm Tr}\,}
\newcommand{\nablaslash}{\nabla\hspace{-.65em}/\hspace{.3em}}
\newcommand{\Aslash}{A\hspace{-.65em}/\hspace{.3em}}
\newcommand{\partialslash}{\partial\hspace{-.5em}/\hspace{.15em}}
\newcommand{\kslash}{k\hspace{-.5em}/\hspace{.15em}}
\newcommand{\llangle}{\left\langle}
\newcommand{\rrangle}{\right\rangle}
\newcommand{\intpsi}{\int {\cal D}\psi^\dagger {\cal D}\psi\,}
\newcommand{\effop}[1]{\mbox{``$#1$''}} 
\begin{document}
%
%
\rightline{RUB-TPII-27/95}
\rightline{PNPI-TH-2076}
\rightline{hep-ph/9510232}
\rightline{September 1995}
\rightline{revised December 1995}
\vspace{.3cm}
\begin{center}
\begin{large}
{\bf Hadronic matrix elements of gluon operators \\
in the instanton vacuum} \\
\end{large}
\vspace{1.4cm}
{\bf D.I.\ Diakonov}$^{\rm 1, 2}$, {\bf M.V.\ Polyakov}$^{\rm 1, 3}$
{\bf and C. Weiss}$^{\rm 4}$ \\
\vspace{0.2cm}
{\em Institut f\"ur Theoretische Physik II \\
Ruhr--Universit\"at Bochum \\
D--44780 Bochum, Germany}
\end{center}
\vspace{1cm}
\begin{abstract}
\noindent
We propose a method to evaluate hadronic matrix elements of QCD gluon
operators in the instanton vacuum. We construct the ground state of the
interacting instanton ensemble for non-zero $\vartheta$--angle using a 
variational principle. A method to study the $\vartheta$--dependence of 
observables on the lattice is suggested. We then derive the effective 
fermion action, which allows to calculate hadronic correlation functions 
in a $1/N_c$--expansion (Nambu--Jona-Lasinio type effective fermion 
theory). Gluon operators are systematically represented as effective 
fermion operators. Physical matrix elements are obtained after integrating 
the correlation functions over fluctuations of the numbers of instantons. 
The influence of the fermion determinant on the topological susceptibility 
is taken into account. Our effective description gives matrix elements 
fully consistent with the trace and $U(1)_A$ anomalies. The approach 
allows to consistently evaluate the nucleon matrix elements of various 
gluon and mixed quark--gluon operators in a chiral soliton picture of 
the nucleon.
\end{abstract}
\vspace{1cm}
PACS: 12.38.Lg, 11.15.Kc, 11.15.Pg, 14.20.Dh \\
Keywords: \parbox[t]{13cm}{non-perturbative methods in QCD, QCD 
vacuum, instantons, nucleon deep--inelastic structure}
\vfill
\rule{5cm}{.15mm}
\\
\noindent
{\footnotesize $^{\rm 1}$ Permanent address: Petersburg Nuclear Physics
Institute, Gatchina, St.\ Petersburg 188350, Russia} \\
{\footnotesize $^{\rm 2}$ E-mail: diakonov@thd.pnpi.spb.ru} \\
{\footnotesize $^{\rm 3}$ E-mail: maxpol@thd.pnpi.spb.ru } \\
{\footnotesize $^{\rm 4}$ E-mail: weiss@hadron.tp2.ruhr-uni-bochum.de}
%
%
%
%
\newpage \tableofcontents
%
%
\newpage
\section{Introduction}
\setcounter{equation}{0}
\renewcommand{\theequation}{\arabic{section}.\arabic{equation}}
It is now widely recognized that instantons play an important, if not
crucial, role in the dynamics of strong interactions. After
their discovery twenty years ago \cite{BPST75}, their relevance
to hadron phenomenology was explored in a number of
pioneering papers \cite{tH76,CDG78,Sh82}, while a quantitative
theory of the instanton medium was still lacking. A treatment of
the interacting instanton ensemble by means of the Feynman
variational principle showed how the instanton medium
stabilizes itself at rather low densities \cite{DP84_1}. In
particular, the ratio of the average size of the instantons to
the average separation between nearest neighbors,
${\bar\rho}/{\bar R}$, was computed as $\simeq 1/3$, the value
suggested from phenomenological considerations \cite{Sh82}. The
instanton vacuum was also studied in direct lattice
experiments, by the so-called cooling procedure\footnote{For a
more complete list of references, see \cite{Lattice_94}.}
\cite{Teper85,ILMSS86,PV88,CDMPV90,CGHN94,MS95}. It was
observed that instantons and antiinstantons ($I$'s and ${\bar
I}$'s for short) are the only non-perturbative gluon
configurations surviving after a sufficient smearing of the
quantum gluon fluctuations. The measured properties of the
$I{\bar I}$ medium, such as the ratio ${\bar\rho}/{\bar R}$,
appear to be close to those computed from the
variational principle \cite{PV88,CGHN94,MS95}.
\par
The success of the instanton vacuum lies in its explanation of
all phenomena related to the dynamical breaking of chiral
symmetry in QCD. The mechanism of chiral symmetry breaking is the
delocalization of the would-be fermion zero modes of individual
instantons \cite{DP84_2,DP86}, leading to a finite fermion
spectral density at eigenvalue zero, proportional to the chiral
condensate, $\langle\bar \psi \psi\rangle$ \cite{BC80}.
Equivalently, one observes that quarks propagating in the instanton
medium acquire a dynamical momentum--dependent mass
\cite{DP86,P89}. A massless pion appears as a collective excitation.
\par
It should be noted that instantons do not lead to confinement \cite{CDG78}. 
However, it has been realized in many ways that it is chiral symmetry 
breaking, and not confinement, which determines the basic characteristics 
of light hadrons (except, probably, for highly excited 
states) \cite{Sh82,DP86,DP86_prep,D87}. This point of view has recently 
received direct support from lattice measurements of correlation functions 
in the cooled vacuum \cite{CGHN94}. Cooling the quantum fluctuations 
eliminates not only the perturbative one--gluon exchange, but also the 
linear confining potential. Nevertheless, the correlation functions of 
various hadronic currents in the cooled vacuum, where only $I$'s and 
${\bar I}$'s are left, appear to be quite similar to those of the 
true ``hot'' vacuum.
\par
Detailed numerical studies of correlation functions of meson and
baryon currents in the instanton medium by Shuryak, Verbaarschot
and Sch\"{a}fer have shown impressive agreement with
phenomenology \cite{Sh93_rev,Sh93}. Correlation functions in the
instanton vacuum can also be computed analytically
\cite{DP86,DP86_prep}.  By integrating first over the instanton
coordinates, one derives an effective fermion action of the form
of a Nambu--Jona-Lasinio model \cite{DP86_prep}. It exhibits a
many--fermion interaction with a specific spin--flavor
structure, first suggested by 't Hooft \cite{tH76}. The
effective fermion theory can be studied systematically in a
$1/N_c$--expansion. In this formulation, chiral symmetry is broken 
by the instanton--induced many--fermionic interactions. 
Correlation functions of meson currents can be computed as fermion 
loop diagrams with a momentum cutoff defined by the average instanton
size, $\bar{\rho}^{-1} \simeq 600\,{\rm MeV}$. The study of correlation 
functions of baryon currents at large
$N_c$ leads to a picture of baryons as chiral solitons --- $N_c$
valence quarks moving in a self--consistent meson field
\cite{DPP88}. This approach to low--energy baryon structure has
been worked out extensively and gives a good account of the
static properties of non--strange and strange baryons
\cite{Goeke94}. Furthermore, by calculating correlation functions of
baryon currents with various weak and electromagnetic current operators,
baryon form factors have been computed \cite{Chr95}, which are 
found to be in excellent agreement with experiment.
\par
In the description of the deep--inelastic structure of the nucleon, one
faces the task of evaluating nucleon matrix elements of certain
operators containing gluon fields \cite{Kodaira95}. There are
essentially two types of such gluonic operators. Operators of twist 2
are related to the moments of the nucleon structure functions. Operators
of higher twist, such as $F_{\mu\nu}^2$, $F_{\mu\nu}\Fdual_{\mu\nu}$, or
$\bar\psi\Fdual_{\mu\nu}\gamma_\nu\psi$ arise in the description of
power corrections. Their evaluation is a challenge for any model of the
QCD vacuum capable of describing the nucleon. Of particular interest is
the nucleon matrix element of $F\Fdual$, the subject of the so--called
proton spin problem\footnote{For a recent lattice estimate of the
nucleon matrix element of the topological charge, see \cite{DiGiacomo94}.}
\cite{Kodaira95}. The nucleon matrix elements have to be extracted from
correlation functions of nucleon currents with the respective gluon
operators. A related problem, which also requires evaluation of matrix
elements of operators involving fermion and gluon fields, is the
estimation of vacuum condensates of ``mixed'' operators like
$\bar\psi F_{\mu\nu}\sigma_{\mu\nu} \psi$, which give corrections in QCD
sum rules \cite{SVZ79}.
\par
The object of this paper is the study of hadronic matrix elements of 
gluon operators in the instanton vacuum. Our framework is the grand
canonical ensemble of interacting $I$'s and $\bar I$'s, to which
fermions are coupled mainly through the zero modes. The fluctuations of 
the number of $I$'s ($N_+$) and ${\bar I}$'s ($N_-$), are essential to 
realize the renormalization properties of QCD, such as the trace anomaly 
and the $U(1)_A$--anomaly. Our strategy will be as follows. We first 
consider the instanton ensemble in gluodynamics and derive the
distribution of the sizes and numbers of pseudoparticles from a
variational principle \cite{DP84_1}. With this instanton size
distribution, we determine the effective fermion action in
quenched approximation, for a fixed number of instantons, $N_\pm$.
Correlation functions are then evaluated in a two--step
procedure. In the first step, we average over instanton
coordinates and the fermion field for fixed $N_\pm$. For correlation
functions of operators consisting of fermion fields (such as meson or 
baryon currents), this average can be performed directly with 
the help of the effective fermion action. Moreover, as we shall
show, even correlation functions with gluon operators can be
reduced to averages in the effective fermion theory: in the
course of integrating over instanton coordinates, the QCD gluon
operator is replaced by an effective many--fermion operator. In
the final step, we then pass from canonical to grand canonical
averages, by integrating over fluctuations of the number of 
$I$'s and $\bar I$'s, $N_\pm$. The probability for these
fluctuations is given by the instanton partition function including
the fermion determinant. By this procedure, we can calculate
correlation functions of nucleon currents with gluon operators,
and extract the nucleon matrix elements by taking appropriate
large--time limits.
\par
Our intention here is twofold. First, we want to set up a framework for 
concrete calculations. All correlation functions are systematically 
reduced to averages in a Nambu--Jona-Lasinio type effective fermion 
theory, which can be evaluated using established techniques \cite{Goeke94}. 
Second, we want to demonstrate that the fundamental renormalization
properties of QCD are preserved in this approach, even at the level of 
hadronic matrix elements. For this reason, we shall focus on the operators 
$F_{\mu\nu}^2$ and $F_{\mu\nu}\Fdual_{\mu\nu}$. In QCD, the nucleon matrix 
element of $F_{\mu\nu}^2$ is uniquely fixed by the trace anomaly, while the 
matrix element of $F_{\mu\nu}\Fdual_{\mu\nu}$ is related to the isoscalar 
axial charge by the $U(1)_A$ anomaly. In the instanton formulation, the 
values of these operators are related, respectively, to $N_+ + N_-$ and 
$N_+ - N_-$. We shall show that the nucleon matrix elements calculated 
according to our prescription are obtained in agreement with the QCD 
anomalies. The proper inclusion of non--topological and topological 
fluctuations of the numbers of instantons will be crucial for this 
property.
\par
Matrix elements of gluon operators depend in general on the 
renormalization point. When computed in the instanton vacuum, they are 
obtained at the scale set by the instanton background, $1/\bar\rho$. In 
general, the results must be subjected to QCD evolution, to make contact 
with experimental data at higher momentum scales. This problem will be 
addressed elsewhere. Here, we concentrate on gluon operators related to 
QCD anomalies, about which exact statements can be made in QCD at the 
hadronic scale.
\par
The plan of this paper is as follows. In section 2, we consider the 
grand canonical ensemble of instantons
in gluodynamics. The distribution of the total number of instantons,
$N_+ + N_-$, is derived from the QCD trace anomaly. We discuss the 
dependence of physical quantities on the $\vartheta$--angle and suggest 
a way to measure this dependence on the lattice. We then perform a 
variational estimate of the grand canonical partition function at finite 
$\vartheta$--angle, allowing for a general form of average pseudoparticle 
interaction with different strengths between same and opposite--kind 
instantons. In section 3, we include fermions in the partition function. 
We derive the effective fermion action in zero mode approximation, for 
an ensemble with a fixed number of instantons, $N_\pm$, but allowing for 
$N_+ \neq N_-$. We outline the bosonization of the 't Hooft interaction 
in the case of more than one fermion flavor ($N_f > 1$). We then discuss 
the suppression of the topological susceptibility due to light fermions 
in QCD on the basis of the $U(1)_A$ and chiral Ward identities. We show 
that the fermion determinant obtained in the instanton vacuum for 
$N_+ \neq N_-$ precisely describes this suppression. In section 4, we 
consider correlation functions of hadronic currents with gluon operators.
We show how fixed--$N_\pm$ correlation functions can be represented as 
averages over the effective fermion theory, in which the gluon operator 
is replaced by an effective fermion operator. We formulate the rules how 
to ``translate'' an arbitrary gluon operator into an effective fermion 
operator, and demonstrate their consistency. The correlation functions of 
the grand canonical ensemble are then obtained by averaging over 
fluctuations of $N_+ + N_-$ and $N_+ - N_-$. It is shown that the basic 
renormalization properties of QCD are preserved in this framework: we 
verify the low--energy theorem of scale invariance for correlation 
functions with the operator $\int d^4 x \, F_{\mu\nu}^2$, and the 
realization of the $U(1)_A$ anomaly for the operator 
$\int d^4 x \, F_{\mu\nu}\Fdual_{\mu\nu}$. In section 5, we then derive 
the nucleon matrix elements of 
$F_{\mu\nu}^2$ and $F_{\mu\nu}\Fdual_{\mu\nu}$ in our approach. The 
nucleon matrix element of $F_{\mu\nu}^2$ is obtained consistent with the 
trace anomaly. The matrix elements of $F_{\mu\nu}\Fdual_{\mu\nu}$ reduces 
to the isosinglet axial coupling constant, $g_A^{(0)}$, as calculated 
within the effective fermion theory. This allows us to consistently 
interpret the result of a former chiral soliton calculation of 
$g_A^{(0)}$ as an estimate of the nucleon matrix element of 
$F_{\mu\nu}\Fdual_{\mu\nu}$ in the large--$N_c$ limit \cite{BlotzPG93}.
Conclusions and an outlook are presented in section 6.
\section{Statistical mechanics of instantons}
\setcounter{equation}{0}
\renewcommand{\theequation}{\arabic{section}.\arabic{equation}}
\subsection{The grand partition function of instantons}
The essence of the instanton vacuum is that the number of 
``pseudoparticles'' is not fixed, hence one is dealing with a {\em grand}
canonical ensemble. Let us denote the number of $I$'s and ${\bar I}$'s,
respectively, by $N_+ , N_-$. The grand partition function is
generically given as
\be
Z (\mu , \vartheta ) &=& \sum_{N_+, N_-}
\exp \left( (\mu + i \vartheta ) N_+ \right)
\exp \left( (\mu - i \vartheta ) N_- \right) Z_{N_\pm} ,
\label{GPF} \\
Z_{N_\pm} &=& \frac{1}{N_+ ! N_- !}
\prod_I^{N_+ + N_-} \int d^4 z_I\, d\rho_I\, dU_I\; d_0(\rho_I)\;
\exp ( -U_{\rm int}) .
\label{CPF}
\ee
Here, the integrals go over the collective coordinates of instantons:
their center coordinates, $z_I$, sizes, $\rho_I$, and orientations, given
by $SU(N_c )$ unitary matrices in the adjoint representation, $U_I$; 
$dU_I$ means the Haar measure normalized to unity. It is useful to 
introduce complex chemical potentials for $I$'s and ${\bar I}$'s, 
$\mu\pm i\vartheta$, where $\vartheta$ is the usual vacuum angle. The 
instanton interaction potential, $U_{\rm int}$, depends on the
separation between pseudoparticles, $z_I - z_J$, their sizes,
$\rho_I, \rho_J$, and their relative orientation, $U_I U_J^\dagger$.
By $d_0(\rho )$ we denote the one--instanton weight. In the one--loop
approximation it is given by \cite{tH76,B79}
\be
d_0 (\rho ) &=& \frac{C_{N_c}}{\rho^5}\beta (M_{\rm cut})^{2N_c}
\exp\left[ -\beta(\rho) \right] ,
\label{d01}
\ee
where $\beta(\rho)$ is the one--loop inverse charge,
\be
\beta (\rho ) &=& \frac{8\pi^2}{g^2(\rho )} \; = \;
b\log\left(\frac{1}{\Lambda\rho}\right) , \hspace{2em}
b \; = \; \frac{11}{3} N_c .
\label{b_glue}
\ee
Note that $\beta$ in the pre-exponential factor of eq.(\ref{d01}) starts 
to ``run'' only at the two--loop level, hence its argument is taken 
at the ultra-violet cut-off, $M_{\rm cut}$. The coefficient $C_{N_c}$ 
depends on the renormalization scheme; in the Pauli--Villars scheme it is
\be
C_{N_c} &=& \frac{4.60 \, {\rm e}^{-1.68 N_c}}{\pi^2 (N_c - 1)! 
(N_c - 2)!}.
\label{CNc}
\ee
The relation between the QCD scale parameters, $\Lambda$, in different
schemes is
\be
\Lambda_{\rm PV} &=& 1.09\, \Lambda_{\overline{\rm MS}} \;\; = \;\;
31.32\, \Lambda_{\rm lattice} .
\ee
If the scheme is changed, one has to change the coefficient according to 
$C_{N_c} \rightarrow C^\prime_{N_c} = C_{N_c}(\Lambda/\Lambda^\prime)^b$.
\par
In the two--loop approximation, the instanton weight is given 
by \cite{DP84_1,VZNS82}
\be
d_0 (\rho ) &=& \frac{C_{N_c}}{\rho^5}\beta(\rho )^{2N_c}
\exp\left[ -\beta_{\rm II}(\rho ) +
\left(2N_c-\frac{b^\prime}{2b}\right)\frac{b^\prime}{2b}
\frac{\log\beta(\rho )}{\beta(\rho )} + O(\beta (\rho )^{-1})
\right] ,
\label{d02}
\ee
where $\beta_{\rm II}(\rho )$ is the inverse charge to two--loop accuracy,
\be
\beta_{\rm II}(\rho ) &=& \beta(\rho)
\, + \, \frac{b^\prime}{2b}\log\frac{2\beta(\rho)}{b},
\hspace{2em} b^\prime \; = \; \frac{34}{3}N_c^2.
\label{2l}
\ee
The partition function, eq.(\ref{GPF}), is actually normalized to the
perturbative partition function (without instantons). Therefore,
when the four--dimensional volume, $V$, goes to infinity, which we always
assume, the logarithm of $Z$ gives the ground state (vacuum) energy
density, $\theta_{44}$, with that of perturbation theory subtracted,
\be
Z &=& \exp\left[-V \left(\theta_{44} - (\theta_{44})_{\rm pert}\right)
\right] .
\label{Fey}
\ee
Here, $\theta_{\mu\nu}$ is the energy--momentum tensor. If the vacuum
state is isotropic, one has $\theta_{44} = \frac{1}{4}\theta_{\mu\mu}$. 
Using the trace anomaly,
\be
\theta_{\mu\mu} &=& \frac{\beta(g^2 )}{4 g^4}
F_{\mu\nu}^2 \; \simeq \; -\frac{b}{32\pi^2} F_{\mu\nu}^2 ,
\label{TA}
\ee
where $F_{\mu\nu}^2 \equiv F_{\mu\nu}^a F_{\mu\nu}^a$ is the gluon field 
strength squared and $\beta(g^2)$ is the Gell-Mann--Low function (not to 
be confused with the inverse charge $\beta(\rho )$),
\be
\beta (g^2 ) &\equiv& \frac{d g^2 (M_{\rm cut} )}{d \log M_{\rm cut}} 
\; = \;
-b\frac{g^4}{8\pi^2} \, - \, \frac{b^\prime}{2}\frac{g^6}{(8\pi^2)^2}
\, - \, \ldots ,
\label{GML}
\ee
one gets \cite{DP84_1}
\be
Z &=& \exp \left( V \frac{b}{4} \frac{1}{32\pi^2}
\llangle F_{\mu\nu}^2 \rrangle_{\rm np} \right) .
\label{GCdef}
\ee
Here, $\langle F_{\mu\nu}^2\rangle_{\rm np}$ is the gluon field vacuum
expectation value, which is due to non-perturbative fluctuations, i.e.
the gluon condensate \cite{NSVZ81}. It is important that this is a
renormalization group--invariant quantity\footnote{To be more precise, 
the RNG--invariant quantity is $\langle\theta_{\mu\mu}\rangle$,
see eq.(\ref{TA}). However, if the coupling constant at the ultra-violet
cutoff scale, $g^2 (M_{\rm cut})$, is small enough, it is sufficient to 
use the beta function to one--loop order.}, meaning that its dependence 
on the ultraviolet cutoff, $M_{\rm cut}$, and the bare charge given at 
this cutoff, $g^2 (M_{\rm cut})$, is such that it is actually 
cutoff--independent,
\be
\frac{1}{32\pi^2} \llangle F_{\mu\nu}^2 \rrangle_{\rm np}
&=&
c \left[ M_{\rm cut} \exp\left(-\int_{...}^{g^2(M_{\rm cut})}
\frac{dg^2}{\beta(g^2 )}\right)\right]^4
\; \simeq \; c^\prime M_{\rm cut}^4 \exp\left[
-\frac{32\pi^2}{bg^2 (M_{\rm cut})} \right] . \nonumber \\
\label{RI}
\ee
(Here, $c, c^\prime$ are constants independent of $M_{\rm cut}$.) By 
definition of the gluodynamics partition function,
\be
Z_{\rm glue} &=& \int \cD A_\mu \exp\left( -\frac{1}{4 g^2 (M_{\rm cut})}
\int d^4 x \, F_{\mu\nu}^2 \right) ,
\label{QCD}
\ee
the l.h.s.\ of eq.(\ref{RI}) is equal to
\be
\frac{1}{32\pi^2} \llangle F_{\mu\nu}^2 \rrangle_{\rm np} &=&
-\frac{1}{V} \frac{1}{8 \pi^2} \frac{d \log Z_{\rm glue}}
{d [1/g^2 (M_{\rm cut})]} .
\ee
Applying the same differentiation to eq.(\ref{RI}) once more, one gets a
low-energy theorem \cite{NSVZ81},
\be
\frac{1}{(8 \pi^2 )^2} \frac{d^2\log Z_{\rm glue}}
{d [1/g^2 (M_{\rm cut})]^2}
&=& \frac{1}{(32\pi^2 )^2} \llangle \int d^4 x \, F_{\mu\nu}^2
\int d^4 y F_{\mu\nu}^2 \rrangle_{\rm np}  -
\frac{1}{(32\pi^2 )^2} \llangle \int d^4 x \, F_{\mu\nu}^2
\rrangle_{\rm np}^2
\nonumber \\
&=& \frac{4}{b} \frac{1}{32\pi^2} \llangle \int d^4 x \, F_{\mu\nu}^2
\rrangle_{\rm np} .
\label{LET}
\ee
If the bare coupling, $g^2(M_{\rm cut})$, is not chosen small enough, 
there are obvious corrections to this formula, following from the 
higher-order terms in the beta function, eq.(\ref{GML}).
\par
This low-energy theorem has important consequence for instantons: it
predicts the dispersion of the number of pseudoparticles in the grand
canonical ensemble \cite{DP84_1}. Assuming the instanton ensemble to be
sufficiently dilute, and taking into account that the one--instanton 
action is
\be
\left( \int d^4 x \, F_{\mu\nu}^2 \right)_{\rm 1-inst} &=& 32\pi^2 ,
\label{1IA}
\ee
one can rewrite the low-energy theorem, eq.(\ref{LET}), as
\be
\langle N^2\rangle - \langle N\rangle^2
&=& \frac{4}{b}\langle N\rangle \;\; = \;\;
\frac{12}{11\, N_c} \langle N\rangle,
\label{NN}
\ee
where $N \equiv N_+ + N_-$ is the total number of pseudoparticles. Thus, 
it follows directly from the renormalization properties of the Yang--Mills 
theory that the dispersion of the number of instantons is less than for a 
free gas of instantons, for which one would get a Poisson distribution,
$\langle N^2\rangle - \langle N\rangle^2 = \langle N\rangle$.
At $N_c \rightarrow \infty$, the dispersion becomes zero.
\par
One concludes that the interaction of instantons with each other is
crucial to support the necessary renormalization properties of the
underlying theory: any cutoff of the integrals over instanton sizes
``by itself'', such as a cutoff due to a possible infrared fixed point 
of the beta function, as suggested recently by Shuryak \cite{Sh95_prep},
leads to the Poisson distribution and hence to the violation of the 
low energy theorem, eq.(\ref{NN}). 
\par
By further differentiating $\log Z_{\rm glue}$ with respect to the bare
coupling, $g^2 (M_{\rm cut})$, one can easily generalize eq.(\ref{NN}) to 
higher moments of the distribution. For example,
\be
\langle N^3\rangle \, - \, 3\langle N^2\rangle\langle N\rangle
\, + \, 2\langle N\rangle^3
&=& \left(\frac{4}{b}\right)^2 \langle N\rangle,
\label{NNN}
\ee
{\em etc.}. Summarizing these formulas, one concludes that the 
distribution of the number of pseudoparticles should, for large 
$\langle N \rangle $, be given by
\be
P(N) &\propto& \exp\left[ -\frac{b}{4} N \left(
\log\frac{N}{\langle N\rangle} -1\right)\right] .
\label{DN}
\ee
A convenient way to generate all moments of the number distribution is
to use the chemical potential, $\mu$, introduced in eq.(\ref{GPF}).
Renormalizability predicts that the grand partition function,
eq.(\ref{GPF}), as a function of $\mu$, should behave as
\be
Z(\mu ) &=& \exp\left[ \frac{b}{4} \langle N\rangle\;
\exp\left( \frac{4\mu}{b} \right) \right] .
\label{Zmu}
\ee
Differentiating $\log Z(\mu)$ with respect to $\mu$ and setting $\mu = 0$,
one obtains all the moments in accordance with the low-energy theorems.
\subsection{$\vartheta$--angle dependence}
The statements made in the previous section are independent of the
$\vartheta$--angle and should thus hold at any $\vartheta$. This means
that in eqs.(\ref{DN}, \ref{Zmu}) only the quantity $\langle N \rangle$ 
may depend on $\vartheta$. Hence the partition function can be written in 
the form
\be
Z(\mu , \vartheta ) &=& \exp\left[ \frac{b}{4}\langle
N\rangle_0 \; f(\vartheta ) \;
\exp \left(\frac{4\mu}{b}\right) \right] , \hspace{2em} f(0) \; = \; 1,
\label{Zmt}
\ee
where $\langle N\rangle_0$ is the average total number of
pseudoparticles at $\vartheta = 0$. The function $f(\vartheta )$ can 
not be determined from general considerations. Nevertheless, comparing
eq.(\ref{Zmt}) and eq.(\ref{GPF}), one realizes that eq.(\ref{Zmt}) 
imposes a non-trivial restriction on the instanton interactions. The 
factorization of the $\mu$-- and $\vartheta$--dependence would be a 
property of non-interacting particles, but in this case the factors $b/4$ 
would be replaced by unity.
\par
Information on $f(\vartheta )$ can be extracted by differentiating 
$\log Z$ with respect to $\vartheta$. The second derivative is 
proportional to the topological susceptibility of the vacuum,
\be
\langle Q_t^2\rangle &\equiv& \llangle \frac{1}{32\pi^2}
\int d^4 x \, F\Fdual \frac{1}{32\pi^2} \int d^4 y \, F\Fdual \rrangle 
\;\; = \;\; -\frac{\partial^2 \log Z}
{\partial \vartheta^2} \nonumber \\
&=& \langle(N_+ - N_-)^2\rangle 
\;\; = \;\; - \frac{b}{4}\langle N\rangle_0
\frac{\partial^2 f(\vartheta )}{\partial \vartheta^2}.
\label{QQ}
\ee
\par
The dependence of the QCD observables on the $\vartheta$ angle is of
fundamental importance. However, until now no direct lattice
measurements with non-zero $\vartheta$ have been performed. The reason is
that the inclusion of a non-zero $\vartheta$--term in lattice
simulations is ambiguous and, moreover, makes the integration measure 
complex. The above discussion suggests an unambigious method to study
the $\vartheta$--dependence of observables on the lattice. Indeed, cooling
down a gluon configuration, one finds --- and that rather early --- the
topological charge of the configuration, $Q_t = \Delta \equiv N_+ - N_-$,
which is the canonical conjugate to $\vartheta$. If $P(\Delta )$ is the
fraction of configurations with given $\Delta$, the probability of a
configuration with given $\vartheta$ is given by the Fourier transform,
\be
Z(\vartheta ) &=& \sum_{\Delta} P(\Delta )\; \exp( i\vartheta\, \Delta ).
\label{Ztheta}
\ee
Measuring a QCD observable, $O$, such as the string tension {\em etc.},
for a set of configurations, and knowing the topological charge of those
configurations, one can establish the dependence of the vacuum average
of the observable on $\vartheta$,
\be
O(\vartheta ) &=& \sum_{\Delta} O(\Delta ) P(\Delta )\;
\exp( i\vartheta \Delta ).
\label{obsthe}
\ee
\par
The topological charge of a configuration, $\Delta = N_+ - N_-$, can 
rather easily be established after cooling. The total number of $I$'s and 
$\bar I$'s, $N = N_+ + N_-$, is a less well--defined quantity, as it tends 
to decrease during the lattice cooling. The question arises: When to stop 
the cooling procedure? When does one hit the true instanton vacuum in the 
lattice simulation? The answer is provided by the dispersion of $N$, 
eq.(\ref{NN}), or, more generally, by eqs.(\ref{DN}, \ref{Zmu}), which 
differ from the Poisson distribution by factors of $b/4 = 11\, N_c/12$. 
When that distribution in the total number of pseudoparticles is achieved, 
one can measure various observables like the condensates, correlation 
functions, {\em etc.}. The most non-trivial check, however, that one is 
dealing with the true instanton vacuum, is the abovementioned factorization 
of the $\mu$-- and $\vartheta$--dependence in the logarithm of the partition 
function. Let the probability of a configuration with $N_+$ $I$'s and $N_-$
$\bar I$'s be $P(N_+, N_-)$. If the original gluon configurations are
cooled sufficiently, but not over-cooled, the grand partition function
should have the form of eq.(\ref{Zmt}),
\be
Z(\mu , \vartheta ) &=& \sum_{N_{\pm}} P(N_+,N_-)\;\exp[\mu(N_++N_-)]\;
\exp[i\vartheta(N_+ - N_-)] \nonumber \\
&=& \exp\left[ \frac{b}{4}\langle N\rangle_0\;f(\vartheta )\;
\exp\left(\frac{4\mu}{b}\right) \right] .
\label{ZmtP}
\ee
Once such non-trivial behavior in $\mu$ and $\vartheta$ is established,
one can read off the gluon condensate as
\be
\frac{1}{32\pi^2} \llangle F_{\mu\nu}^2 \rrangle_{\rm np}
&=& \frac{\langle N\rangle_0}{V}.
\label{GC}
\ee
\par
The above statements are about all one can say on the dispersion
of pseudoparticles on general grounds. To obtain more information on
the instanton vacuum, one has to go into dynamics, {\em i.e.}, take
into account the instanton--instanton interaction in a quantitative
way. In the next section, we estimate the ground state of the
grand canonical partition function, eq.(\ref{GPF}), using a
variational principle.
\subsection{Variational estimate of the grand partition function}
In this section, we show that actually very limited information on
the instanton interactions is needed to make a variational estimate
of the basic properties of the instanton ensemble. We shall neglect the
many-body forces, an approximation which is justified
{\em a posteriori}, if the medium turns out to be sufficiently dilute.
\par
An intuitively clear description is achieved by introducing the actual 
size distributions of instantons in the medium, which do not necessarily 
coincide with the weight of an isolated
instanton, $d_0(\rho)$, of eqs.(\ref{d01}, \ref{d02}). Moreover,
the results of section 2.1 show that they necessarily do not coincide 
with $d_0 (\rho)$ --- otherwise the renormalization properties of the 
theory would be violated. Following ref.\cite{DP84_1}, we now derive 
the ``best'' distribution functions, $d_\pm (\rho)$, from a variational 
principle. This approach preserves all the general requirements discussed 
in sections 2.1 and 2.2.
\par
As an intermediate step, it will be convenient to work with the canonical
ensemble. We first derive the best $d_\pm (\rho)$ for a given number of
$I$'s and $\bar I$'s, and leave the summation over $N_\pm$ for the end.
Let us write the partition function for given $N_\pm$, eq.(\ref{CPF}),
in the form
\be
Z_{N_\pm} &=& \frac{1}{N_+! N_-!} \int \prod_I^{N_+ + N_-} d\xi_I
\exp [ -E(\xi )], \hspace{2em} E(\xi ) \; = \;
-\sum_I^{N_+ + N_-} \log d_0(\rho_I ) \, + \, U_{\rm int}(\xi ) .
\nonumber \\
\label{ZNN}
\ee
Here, $\xi_I = (z_I , \rho_I , U_I )$ denote the set of collective 
coordinates of the $I$th pseudoparticle, and $\xi$ those of all 
$N_+ + N_-$ pseudoparticles.
\par
For the purpose of a variational estimate, we introduce a
``non-interacting'' partition function,
\be
Z_1 &=& \frac{1}{N_+! N_-!}\int \prod_I^{N_+} d\xi_I\, d_+(\rho_I )
\prod_{\bar I}^{N_-} d\xi_{\bar I}\, d_-(\rho_{\bar I} )
\; = \; \frac{1}{N_+! N_-!}\int \prod_I^{N_+ + N_-}
d\xi_I \exp[-E_1(\xi )], \nonumber \\
E_1 (\xi) &=& -\sum_I^{N_+} \log d_+(\rho_I)
-\sum_{\bar I}^{N_-} \log d_-(\rho_{\bar I}).
\label{Z1}
\ee
Here, $d_\pm(\rho)$ are the effective size distributions in the
ensemble of interacting $N_+$ $I$'s and $N_-$ $\bar I$'s. Note that these
distributions, naturally, depend on $N_\pm$. The partition function,
eq.(\ref{Z1}), is immediately expressed through fugacities, $\zeta_\pm$,
\be
Z_1 &=& \frac{1}{N_+! N_-!}(V\zeta_+)^{N_+}(V\zeta_-)^{N_-},
\hspace{2em} \zeta_\pm \; = \; \int d\rho\, d_\pm (\rho ) .
\label{fugac}
\ee
To find the best $d_\pm(\rho )$ for a given interaction potential,
$U_{\rm int}$, we use the Feynman variational principle, which is based 
on the convexity of the exponential function. It states that
\be
Z_{N_\pm} &=& Z_1 \, \overline{\exp[-(E - E_1 )]}
\;\; \ge \;\; Z_1 \exp[ - \overline{(E - E_1 )} ] .
\label{Feyn}
\ee
Here the averaging is performed with the modified partition function, 
$Z_1$,
\be
\overline{A(\xi )} &\equiv& \frac{1}{Z_1}
\frac{1}{N_+! N_-!} \int \prod_I^{N_+} d\xi_I \, d_+ (\rho_I )
\prod_{\bar I}^{N_-} d\xi_{\bar I} \, d_- (\rho_{\bar I}) \, A(\xi ).
\label{avdef}
\ee
In our case we have \cite{DP84_1}
\be
\overline{E - E_1} &=&
\sum_{\epsilon = \pm} \frac{N_\epsilon}{\zeta_\epsilon}\int d\rho
\log\frac{d_\epsilon(\rho)}{d_0(\rho)} \nonumber \\
&+& \frac{1}{2V^2}\sum_{\epsilon_1, \epsilon_2 = \pm}
\frac{N_{\epsilon_1}}{\zeta_{\epsilon_1}}
\frac{N_{\epsilon_2}}{\zeta_{\epsilon_2}}\int d\xi_1 d\xi_2 \;
d_{\epsilon_1}(\rho_1) \, d_{\epsilon_2}(\rho_2) \,
U_{\rm int} (\xi_1, \xi_2 ).
\label{EE1}
\ee
The best effective one--particle distributions, $d_\pm(\rho)$, are those 
which maximize the r.h.s.\ of the inequality eq.(\ref{Feyn}), {\em i.e.}, 
those which reproduce best the true partition function. We see from 
eq.(\ref{EE1}), that in this approach the optimum distribution depends 
only on averages of the interaction potential over separations and 
orientations. There are two basic quantities, the average interaction 
between instantons of the same and of opposite kind; different average 
interaction between $II$ and $\bar I \bar I$ are prohibited by $CP$ 
invariance. Let us denote the basic averages by
\be
\int dU_1 \, dU_2 \, d^4 z_1 \, d^4 z_2 \; 
U_{\rm int}^{I\bar I \, (\bar I I)}
(z_1 - z_2, \rho_1, \rho_2, U_1 U_2^\dagger )
&=& V\beta \rho_1^2 \rho_2^2 \gamma_a ,
\label{gammaa} \\
\int dU_1 \, dU_2 \, d^4 z_1 \, d^4 z_2 \; 
U_{\rm int}^{II \, (\bar I\bar I)}
(z_1 - z_2, \rho_1, \rho_2, U_1 U_2^\dagger )
&=& V\beta \rho_1^2 \rho_2^2 \gamma_s .
\label{gammas}
\ee
\par
Some comments are in order here. The factors $V$ and $\beta$ on the 
r.h.s.\ of eqs.(\ref{gammaa}, \ref{gammas}) are trivial consequences of 
the fact that the potential depends only on the separation, and that it 
is proportional to the inverse charge, $\beta$. Less trivial is, which 
argument one should one choose for the running $\beta$. Following the 
reasoning of \cite{DP84_1}, we are inclined to write 
$\beta (\langle R \rangle )$, where $\langle R \rangle$ is the average 
separation, to be determined below. Fortunately, this question is not very 
important quantitatively, as the dependence on the argument is only 
logarithmic (see also the discussion of the numerical values below). The 
product of $\rho_1^2 \rho_2^2$ arises because of dimensions, and because 
the interaction has to vanish at zero sizes. Finally, we have introduced 
here two phenomenological constants for the average interaction of 
pseudoparticles of the same and opposite kind, $\gamma_s$ and $\gamma_a$. 
It should be mentioned that for the sum ansatz one has \cite{DP84_1}
\be
\gamma_a \; = \; \gamma_s &=& \frac{27}{4}\pi^2\frac{N_c}{N_c^2-1} .
\label{sumanz}
\ee
However, here we would like to take a more liberal stand and allow
arbitrary $\gamma_s , \gamma_a$. Nevertheless, the behavior
of $\gamma_s , \gamma_a$ at large $N_c$ as $1/N_c$, shown by
eq.(\ref{sumanz}), is probably a general statement --- it arises due to
averaging over relative orientations in eqs.(\ref{gammaa}, \ref{gammas}).
\par
Substituting the definitions eqs.(\ref{gammaa}, \ref{gammas}) into
eqs.(\ref{fugac}, \ref{Feyn}, \ref{EE1}), we get
\be
Z_{N_\pm} &\ge& \frac{1}{N_+! N_-!} \exp\left(
\sum_{\epsilon = \pm} N_\epsilon \left[\log(V\zeta_\epsilon)-
\overline{\log\frac{d_\epsilon(\rho)}{d_0(\rho)}}\right] 
\right) \nonumber \\
&& \times\exp\left(-\frac{\beta}{2V}
\sum_{\epsilon_1, \epsilon_2 = \pm}N_{\epsilon_1}N_{\epsilon_2}
\overline{\rho_{\epsilon_1}^2}\overline{\rho_{\epsilon_2}^2}
\gamma_{\epsilon_1\epsilon_2}\right) .
\label{Zvar}
\ee
Here, we have introduced the average sizes,
\be
\overline{\rho_\pm^2} &=& \frac{1}{\zeta_\pm}\int d\rho \,
d_\pm(\rho) \rho^2.
\label{rhoav}
\ee
Varying eq.(\ref{Zvar}) with respect to $d_\pm(\rho )$ we find the best
effective size distribution, which takes into account the interactions
with the medium,
\be
d_\pm(\rho) &=& C_\pm \, d_0(\rho) \, \exp(-\rho^2\alpha_\pm ), 
\label{bestd} \\
\alpha_+ &=& \frac{\beta}{V}\left(\gamma_s\overline{\rho_+^2}N_+
+\gamma_a\overline{\rho_-^2}N_-\right) ,
\hspace{2em}
\alpha_- \; = \; \frac{\beta}{V}\left(\gamma_a\overline{\rho_+^2}N_+
+\gamma_s\overline{\rho_-^2}N_-\right).
\ee
The coefficients $C_\pm$ cancel in eq.(\ref{Zvar}) and can be both put
to unity.
\par
We see that the one--instanton size distributions are multiplied by
gaussian cutoff functions, whose slope depends on the density of
particles. Substituting eq.(\ref{bestd}) into 
eqs.(\ref{fugac}, \ref{rhoav}) one obtains the fugacities, $\zeta_\pm$, 
and the average sizes of the instantons, $\overline{\rho_\pm^2}$. In this 
way, we get the best variational approximation to the fixed--$N_\pm$ 
partition function,
\be
Z_{N_\pm} &\ge& \frac{1}{N_+! N_-!} \exp \left(
\sum_{\epsilon = \pm} N_\epsilon \log (V \zeta_\epsilon )
+ \frac{\beta}{2V}
\sum_{\epsilon_1, \epsilon_2 = \pm} N_{\epsilon_1} N_{\epsilon_2}
\overline{\rho_{\epsilon_1}^2} \overline{\rho_{\epsilon_2}^2}
\gamma_{\epsilon_1\epsilon_2} \right) .
\label{bestZ}
\ee
\par
Taking the one--loop expression for $d_0(\rho )$, eq.(\ref{d01}), one
finds
\be
\zeta_\pm &=& \int d\rho \, d_0(\rho) \exp ( -\rho^2\alpha_\pm )
\;\; = \;\; \frac{1}{2}C_{N_c}{\tilde \beta}^{2N_c} \Lambda^b
\Gamma(\nu)\alpha_\pm^{-\nu},
\ee
\be
\overline{\rho_\pm^2} &=& -\frac{\partial \log \zeta_\pm}{\partial
\alpha_\pm} \; = \; \frac{\nu}{\alpha_\pm} , \hspace{2em}
\nu \; = \; \frac{b-4}{2} , \hspace{2em} \tilde{\beta} \; = \; 
\beta (M_{\rm cut}) .
\label{zeta1}
\ee
We get a system of equations relating the average sizes of $I$'s and
$\bar I$'s to their densities,
\be
\overline{\rho_+^2} \, \left( \gamma_s \,\overline{\rho_+^2}\,
\frac{N_+}{V} \; + \; \gamma_a \, \overline{\rho_-^2} \, 
\frac{N_-}{V}\right) &=& \frac{\nu}{\beta} , \nonumber \\
\overline{\rho_-^2} \, \left( \gamma_a \, \overline{\rho_+^2}\,
\frac{N_+}{V} \; + \; \gamma_s \overline{\rho_-^2}\frac{N_-}{V}\right) 
&=& \frac{\nu}{\beta}.
\label{selfcons}
\ee
At $N_+ = N_- = N/2$, which will be the case at $\vartheta = 0$, when the 
vacuum is $CP$ symmetric, one has the packing fraction,
\be
\left(\overline{\rho_+^2}\right)^2\frac{N}{V} &=&
\left(\overline{\rho_-^2}\right)^2\frac{N}{V} \; = \;
\frac{\nu}{\beta}\frac{2}{\gamma_s+\gamma_a} .
\label{selfcons0}
\ee
\par
To get the variational estimate of the fixed--$N_\pm$
partition function, $Z_{N_\pm}$,
one has to substitute $\overline{\rho_\pm^2}$ from eq.(\ref{selfcons})
into eq.(\ref{bestZ}).
Our aim is, however, the grand partition function, $Z(\mu , \vartheta )$,
which is obtained from $Z_{N_\pm}$ according to eq.(\ref{GPF}). The
summation over $N_\pm$ in eq.(\ref{GPF}) can, of course, be done by the
steepest descent method. The saddle point values of $N_+$ and $N_-$
are in general complex and conjugate to each other, and both are 
proportional to the four--dimensional volume, $V \rightarrow \infty$. At 
$\vartheta = 0$, the saddle point values of $N_\pm$ coincide, and are real. 
After some straightforward algebra we obtain the logarithm of the grand 
partition function,
\be
\log Z(\mu , \vartheta ) &=& \frac{b}{4}\langle N\rangle_0\;
\exp \left( \frac{4\mu}{b} \right) f(\vartheta ) .
\label{Zmutvar}
\ee
Here, $\langle N\rangle_0$, the average total number of $I$'s and $\bar
I$'s at $\vartheta = 0$ and $\mu = 0$, is given by
\be
\langle N\rangle_0 &=& V\Lambda^4\left[C_{N_c}{\tilde \beta}^{2N_c}
\Gamma\left(\frac{b-4}{2}\right)\right]^{\frac{4}{b}}
\left[\beta\frac{b-4}{2}\frac{\gamma_s+\gamma_a}{2}\right]^
{\frac{4-b}{b}}.
\label{Naver1}
\ee
Note that we have obtained the grand partition function exactly in the
form predicted in section 2 by considering the renormalization properties
of the Yang--Mills theory. Here, we get a concrete formula for the density
of instantons in terms of the $\Lambda_{\rm QCD}$ parameter, and a concrete
form of $f(\vartheta )$. The function $f(\vartheta )$ in eq.(\ref{Zmutvar}) 
is given in an indirect way. One has to first solve the equation 
determining the phase of the saddle point of $N_\pm$, 
$\psi = \log (N_+ / N_- )/2i$,
\be
\psi \; + \; \frac{b-4}{4}\arctan\frac{(1-r^2)\sin\psi}
{\cos\psi+r\sqrt{1-r^2\sin^2\psi}} &=& \vartheta , \hspace{2em}
r \; \equiv \; \frac{\gamma_a}{\gamma_s} ,
\label{phase}
\ee
substitute its solution to determine the saddle point value of the
partition function, and then read off $f(\vartheta )$,
\be
f(\vartheta ) &=& 
\cos\psi\left(\frac{r\cos\psi + \sqrt{1-r^2\sin^2\psi}}{1+r}
\right)^{\frac{4-b}{b}}.
\label{ft}
\ee
This function depends on the ratio, $r$, of the average interactions of
same and opposite kind pseudoparticles, eqs.(\ref{gammaa}, \ref{gammas}).
\par
It should be noted that not all values of $\gamma_{s,a}$ lead to a stable
ensemble. First, $\gamma_s + \gamma_a$ must be positive --- otherwise the
system would collapse. This can be seen from
eqs.(\ref{selfcons}, \ref{Naver1}). Second, $\gamma_s$ itself must be 
positive. If the interaction between same--kind pseudoparticles were on 
the average attractive, the system would prefer to break up into two 
subsystems, with only $I$'s in one part of the universe and only 
$\bar I$'s in the other, with an apparent maximum violation of $CP$ 
symmetry in both parts. Third, the topological susceptibility of the
vacuum, eq.(\ref{QQ}), must be positive. Determining the second derivative
of $f(\vartheta )$ from eqs.(\ref{phase}, \ref{ft}), we find
\be
\langle Q_t^2\rangle &=& \frac{4}{b-r(b-4)} \langle N\rangle_0 .
\label{TS}
\ee
Hence we must require $r < b/(b - 4)$, meaning that the effective 
repulsion of opposite--kind pseudoparticles can not be much stronger than 
that of same--kind.
\par
In several simple cases, the function $f(\vartheta )$ can be found
analytically. For example, at $r = 1$, {\em i.e.}, $\gamma_a = \gamma_s$,
which is the case of the sum ansatz of \cite{DP84_1}, we have
\be
f(\vartheta )_{r = 1} &=& (\cos\vartheta )^{\frac{4}{b}} , 
\label{ft1}
\ee
and the topological susceptibility is
\be
\langle Q_t^2\rangle &=& \langle N\rangle_0 .
\label{QQ_sum}
\ee
In the case $r = 0$ (no effective repulsion between opposite--kind
instantons), we have
\be
f(\vartheta )_{r = 0} &=& \cos\frac{4\vartheta}{b}.
\label{ft2}
\ee
At $r\rightarrow -1$, which is the edge of the stability region, we get
\be
f(\vartheta )_{r\rightarrow -1} &=&
\left[\cos\frac{2\vartheta}{b-2}\right]^{\frac{2(b-2)}{b}}.
\label{ft3}
\ee
By measuring the $\vartheta$--dependence of the partition function
by lattice cooling, as described in section 2, one should be
able to ``experimentally'' determine the ratio $r$. This quantity is
one of the basic characteristics of the instanton vacuum.
\par
To end this section, we present numerical values of the
basic characteristics of the instanton vacuum, which were calculated
using the two--loop instanton density, eq.(\ref{d02}).
One first determines the average instanton sizes from 
eqs.(\ref{rhoav}, \ref{selfcons}). The average instanton density
then follows from maximizing eq.(\ref{bestZ}).
Since the argument of the inverse charge, $\beta$, is not uniquely 
determined from first principles, we set 
$\beta = b \log\left[ a (N/V\Lambda^4)^{1/4} \right]$, with $a$
a dimensionless parameter, and explore different choices of $a$.
Results for the instanton density ({\em i.e.}, the gluon condensate),
$(N/V\Lambda^4 )^{1/4}$, the packing fraction, 
$(\overline{\rho^2}^2 N/V)^{1/4}$, and the effective inverse
charge are shown in table 1. Given there are the results for
two different values of the constants defining the
average instanton interactions, $\gamma_{a, s}$. Here, $N_+ = N_-$,
so that the results depend only on the sum, $\gamma_a + \gamma_s$.
From table 1 one sees that the instanton density and packing fraction
depend only very weakly on the argument of the
inverse coupling, $a$, although the values of $\beta$ vary 
considerably. The values obtained for the density and packing fraction
are rather close to those of \cite{DP84_1}.
%
%
\noindent
\begin{table}[t]
\centering
\[
\begin{array}{|l|c|c|c|c|}
\hline
\rule[-1cm]{0cm}{2cm}
& a & {\displaystyle \left(\frac{N}{V\Lambda^4}\right)^{1/4} }
& {\displaystyle \left(\overline{\rho^2}^2\frac{N}{V}\right)^{1/4} }
& {\displaystyle \beta 
= b \log\left[ a \left(\frac{N}{V\Lambda^4}\right)^{1/4} \right]}
\\
\hline
{\rm I} 
&  2.5 &  .68  &  .37  &  5.8 \\
&  3   &  .66  &  .35  &  7.5 \\
&  3.5 &  .64  &  .34  &  9.0 \\
&  4   &  .63  &  .33  & 10.2 \\
&  5   &  .62  &  .31  & 12.4 \\
&  7   &  .60  &  .30  & 15.8 \\
\hline
{\rm II}
&  2.5 &  .74  &  .42  &  6.8 \\
&  3   &  .72  &  .40  &  8.5 \\
&  3.5 &  .71  &  .39  & 10.0 \\
&  4   &  .70  &  .38  & 11.3 \\
&  5   &  .68  &  .36  & 13.5 \\
&  7   &  .66  &  .34  & 16.9 \\
\hline 
\end{array}
\]
\caption[]{\it The dimensionless instanton density, 
$(N/V\Lambda^4 )^{1/4}$, the packing fraction, 
$(\overline{\rho^2}^2 N/V)^{1/4}$, and the effective inverse
charge, $\beta$, for various values of a parameter $a$,
which defines the argument of the inverse charge 
according to $\beta = b \log\left[ a (N/V\Lambda^4)^{1/4} \right]$.
The two sets correspond to different values of the constant
$\half (\gamma_a + \gamma_s )$ determining the average instanton 
interaction, cf.\ {\rm eqs.(\ref{gammaa}, \ref{gammas})}. 
Set {\rm I} is obtained with 
the sum ansatz, {\rm eq.(\ref{sumanz})}, set {\rm II} with 
$\half(\gamma_a + \gamma_s )$ equal to half the value of the 
sum ansatz.}
\end{table}
\section{Including fermions}
\setcounter{equation}{0}
\renewcommand{\theequation}{\arabic{section}.\arabic{equation}}
\subsection{The effective fermion action}
So far we have discussed the partition function of instantons
in pure gluodynamics. When fermions are included ($N_f$ flavors), the
coefficient of the beta function, eq.(\ref{b_glue}), is changed to
\be
b &=& \frac{11}{3} N_c - \frac{2}{3} N_f .
\label{b_fermions}
\ee
The fixed--$N_\pm$ partition function with fermions, normalized to the
perturbative partition function, is given as
\be
Z^{\,\rm fermions}_{N_\pm} &=& \frac{1}{N_+ ! N_- !}
\int \prod_I^{N_+ + N_-} d\xi_I \, d_0 (\rho_I ) \exp ( -U_{\rm int})
\, {\rm Det}(m, M_{\rm cut}) \label{Z_f_N} ,
\\
{\rm Det}(m, M_{\rm cut}) &\equiv& \frac{\det (i\nablaslash (\xi ) + i m)}
{\det (i\partialslash + i m) } \;
\frac{\det (i\partialslash + i M_{\rm cut})}
{\det (i\nablaslash (\xi ) + i M_{\rm cut})} .
\label{Det_normalized}
\ee
Here, $m = {\rm diag}(m_1 , \ldots , m_{N_f})$ is the bare quark mass 
matrix, $M_{\rm cut}$ the Pauli--Villars regulator mass. Furthermore, 
$\nablaslash (\xi )$ denotes the fermion Dirac operator in the background 
of the $N_\pm$--instanton configuration,
\be
\nablaslash (\xi ) &=& \partialslash - i \Aslash (x; \xi ) , \nonumber \\
A(x; \xi ) &=& \sum_I^{N_+} A_+ (x; \xi_I ) + \sum_{\bar I}^{N_-}
A_- (x, \xi_{\bar I}) .
\label{A_inst}
\ee
(Here, $A_\pm$ are the fields of individual $I$'s and $\bar I's$,
in singular gauge \cite{DP84_1}.)
The eigenvalues of the Dirac operator contributing to the determinant
can can be divided into ``low'' and ``high'' frequencies by introducing
a splitting mass parameter, $M_1$,
\be
{\rm Det} (m, M_{\rm cut}) &=& {\rm Det} (m , M_1 )\, 
{\rm Det} (M_1 , M_{\rm cut}) .
\label{Det_split}
\ee
The value of $M_1$ will be chosen of order $1/\bar{\rho}$. The essence
of the instanton description of chiral symmetry breaking is to treat the 
two factors in eq.(\ref{Det_split}) in different ways \cite{DP84_2,DP86}. 
The high--frequency part can approximately be factorized in one--instanton 
contributions, and is therefore included in the one--instanton weight of 
eq.(\ref{Z_f_N}). The low--frequency part, which is dominated by the 
fermionic zero modes associated with the instantons, must be averaged in 
the background of all $N_\pm$ instantons simultaneously. This leads to the 
delocalization of the zero modes, which is the mechanism of chiral symmetry 
breaking.
\par
When the fermion determinant is included in the variational description
of the interacting instanton partition function, it modifies the effective
instanton size distribution in the medium (in addition to the change in the
$\beta$--function, eq.(\ref{b_fermions})). This modification is of the 
order $N_f / N_c$ \cite{DP86}. To lowest order in this parameter (quenched 
approximation), one gets back the unmodified size
distribution of gluodynamics, eq.(\ref{bestd}). In the following we use 
the gluodynamics size distribution in one--loop approximation to average 
the low--momentum fermion determinant. With the sum ansatz, 
eq.(\ref{sumanz}), the effective size distributions for $I$'s and 
$\bar I$'s coincide, and eq.(\ref{bestd}), with eq.(\ref{d01}), becomes
\be
d_\pm (\rho ) &=& \mbox{\rm const}\times
\rho^{b - 5} \exp\left[-\frac{b - 4}{2}\, 
\frac{\rho^2}{\overline{\rho^2}}\right] ,
\hspace{2em} \overline{\rho^2} \; = \; \overline{\rho_+^2} 
\; = \; \overline{\rho_-^2} .
\label{d_rho}
\ee
Since $b = O(N_c )$, the normalized size distribution reduces to a 
$\delta$--function in the limit $N_c \rightarrow \infty$,
\be
\left( \int d\rho_{\pm}\, d_\pm (\rho) \right)^{-1} d (\rho )
&\rightarrow & \delta (\rho - \bar\rho ) \hspace{2em}
(b \,\rightarrow \, \infty ),
\hspace{2em} \bar\rho \; = \;  \sqrt{\overline{\rho^2}} .
\label{d_norm}
\ee
For simplicity, we shall use the
$\delta$--function distribution in the following calculations\footnote{This
approximation is only a practical simplification, and does not imply that 
we take $b$ to infinity elsewhere. Corrections for finite--width size
distributions can readily be included.}, {\em i.e.}, we replace
all instanton sizes by $\bar\rho$.
\par
The variational approximation to the fixed--$N_\pm$ partition function with
fermions can then be written as
\be
Z^{\,\rm fermions}_{N_\pm} &=& 
Z_{N_\pm} \, \overline{\rm Det}_{N_\pm} , \hspace{2em}
\overline{\rm Det}_{N_\pm} \; \equiv \; \int \prod_I^{N_+ + N_-}
\frac{d^4 z_I}{V} dU_I \, {\rm Det}(m, M_1) .
\label{Z_f_opt}
\ee
Here, $Z_{N_\pm}$ is the variational partition function of gluodynamics. 
We remark that corrections to this factorized form of the partition 
function can in principle be calculated using the variational principle.
\par
In the ground state described by the partition function, 
eq.(\ref{Z_f_opt}), the chiral symmetry of the fermions is spontaneously 
broken \cite{DP84_2,DP86}. A useful concept is the effective fermion 
action \cite{DP86_prep}. It is obtained by performing in eq.(\ref{Z_f_opt}) 
the integral over the instanton coordinates, and is valid in the vicinity 
of the saddle point of eq.(\ref{Z_f_opt}). In terms of the effective action, 
the fixed--$N_\pm$ partition function represented as
\be
\overline{\rm Det}_{N_\pm}
&=& C \intpsi \exp \left( -S_{\rm eff}[\psi^\dagger , \psi ] \right) .
\label{Z_eff}
\ee
(Here, $C$ is a normalization constant.) The effective action can be used 
to calculate averages of operators consisting of fermion 
fields (see section 4).
\par
To explicitly construct the effective fermion action, we need
to know the fermion propagator in the background of the
$N_\pm$--instanton configuration, eq.(\ref{A_inst}). For a single
instanton, the fermion propagator is singular in the chiral limit due
to the zero mode. It may be approximated as the sum of the free propagator
and the explicit contribution of the zero mode \cite{DP86,DP86_prep},
\be
\left(i\nablaslash (\xi_{I(\bar I)} ) + im \right)^{-1}_{\rm 1-inst} 
&\approx&
(i\partialslash )^{-1} \;\; - \;\; \frac{\Phi_\pm (x; \xi_{I(\bar I)} )
\Phi_\pm^\dagger (y; \xi_{I(\bar I)} )}{im} .
\label{prop_zero_mode}
\ee
Here, $\Phi_\pm (x ; \xi_{I(\bar I)})$ is the wave function of the fermion 
zero mode in the background of one $I (\bar I )$.
This interpolating formula should be accurate both at small momenta
($p \ll 1/\bar\rho$), where the zero mode is dominant, and 
at large momenta ($p \gg 1/\bar\rho$), where the propagator reduces to the 
free one. Eq.(\ref{prop_zero_mode}) is equivalent to using an approximate
fermion action \cite{DP86_prep},
\be
\exp \left( -\widetilde{S}^{I(\bar I)}[\psi^\dagger , \psi ] \right) 
&\propto& \exp \left( \sum_f^{N_f} \int d^4 x \, 
\psi^\dagger_f i \partialslash \psi_f \right)
\prod_{f = 1}^{N_f} 
\left( im_f - V^{I(\bar I)}_\pm [\psi^\dagger_f , \psi_f ] \right) ,
\ee
\be
V^{I(\bar I)}_\pm [\psi^\dagger_f , \psi_f ] &=&
\int d^4 x \left( \psi^\dagger_f (x) i\partialslash 
\Phi_\pm (x; \xi_{I(\bar I)} ) \right)
\int d^4 y \left( \Phi_\pm^\dagger (y; \xi_{I(\bar I)} ) i\partialslash
\psi_f (y) \right) .
\label{V_I}
\ee
The fermion Green function calculated with this action coincides with 
eq.(\ref{prop_zero_mode}). Given the diluteness of the instanton medium, 
one may thus approximate the fermion action in the background of an 
$N_\pm$--instanton configuration by the product
\be
\exp\left( -\widetilde{S}[\psi^\dagger , \psi ] \right) 
&\propto&
\exp \left( \sum_f^{N_f} \int d^4 x \, \psi^\dagger_f
i \partialslash \psi_f \right) 
\label{S_zero_mode} \\
&& \times \left( \prod_I^{N_+} \prod_{f = 1}^{N_f}
\left( im_f - V^I_\pm [\psi^\dagger_f , \psi_f ] \right) \right)
\left( \prod_{\bar I}^{N_-} \prod_{f = 1}^{N_f}
\left( im_f - V^{\bar I}_\pm [\psi^\dagger_f , \psi_f ] \right) \right) .
\nonumber
\ee
\par
With the fermion propagator approximated by eq.(\ref{S_zero_mode}), the
average over the configurations of the independent instantons, 
eq.(\ref{Z_f_opt}), reduces to a product of averages over individual $I$'s 
and $\bar I$'s and can be carried out explicitly. Since the splitting 
mass, $M_1$, is of the order of the effective momentum cutoff in the
zero mode wave function, $1/\bar\rho$, the normalizing determinant
of eq.(\ref{Det_normalized}) in the zero mode approximation may simply be 
replaced by $M_1$ to the power of the number of zero modes,
\be
\det (i\nablaslash + i M_1 ) &\propto& (i M_1 )^{N_f (N_+ + N_-)} .
\ee
Its only role in the following will be to make the partition function 
dimensionless. The averaged fermion determinant, eq.(\ref{Z_f_opt}), 
becomes
\be
\overline{\rm Det}_{N_\pm} &=& (i M_1 )^{-N_f (N_+ + N_-)}
\intpsi \int \prod_I^{N_+ + N_-} \frac{d^4 z_I}{V} dU_I \,
\exp\left( -\widetilde{S}[\psi^\dagger , \psi ] \right) \nonumber \\
&=& (i M_1 )^{-N_f (N_+ + N_-)}
\intpsi \exp \left( \sum_f^{N_f} \int d^4 x \, 
\psi^\dagger_f i\partialslash \psi_f \right) \, W_+^{N_+} \, W_-^{N_-} .
\label{S_average}
\ee
Here, $W_\pm$ denote the one--instanton averages
\be
W_\pm [\psi^\dagger , \psi ] &\equiv& 
\int \frac{d^4 z_{I(\bar I)}}{V} dU_{I(\bar I)}
\prod_f^{N_f} \left( im_f - V^{I(\bar I)}_\pm [\psi^\dagger_f , \psi_f ] 
\right) .
\label{A}
\ee
Performing the average over color to leading order in $1/N_c$,
eq.(\ref{A}) can be expressed as\footnote{Explicit expressions
for the zero mode vertex, eq.(\ref{V_I}) and formulas for
integrals over the color orientation matrices can be found in 
\cite{DP86,DP86_prep}.}
\be
W_\pm &=& \left( - \frac{4\pi^2\bar\rho^2}{N_c} \right)^{N_f}
\int \frac{d^4 z}{V} \det J_\pm (z) 
\;\; + \;\; \sum_{f = 1}^{N_f} 
i m_f \left( - \frac{4\pi^2\bar\rho^2}{N_c} \right)^{N_f - 1}
\int \frac{d^4 z}{V} \, {\textstyle{\det'_f}} J_\pm (z)
\nonumber \\
\label{A_det_1} \\
&=& i^{N_f} \left( \frac{4\pi^2\bar\rho^2}{N_c} \right)^{N_f} 
\int \frac{d^4 z}{V} 
\det \left[ i J_\pm (z) + \frac{m N_c}{4\pi^2\bar\rho^2}\right]
\;\; + \;\; O(m^2 ) .
\label{A_det_2}
\ee
Here the determinants are over flavor indices, and
$\det'_f$ denotes the minor in which the $f$--th row and column are
omitted. The currents, $J_\pm (z)$, are color singlets and
$N_f \times N_f$--matrices in flavor,
\be
J_\pm (x)_{fg} &=& \fourint{k}\fourint{l} \exp ( -i(k - l)\!\cdot\! x)
\, F(k) F(l) \, \psi^\dagger_f (k) \half (1 \pm \gamma_5 ) \psi_g (l) ,
\label{J}
\ee
where $F(k)$ is a form factor related to the wave function of the
zero mode in momentum space,
\be
F(k) &=& - t \frac{d}{dt} \left[ I_0 (t) K_0 (t) - I_1 (t) K_1 (t)
\right] \; \rightarrow
\left\{ \ba{ll} 1 & t \rightarrow 0 \\  \textfrac{3}{4} t^{-3}
& t \rightarrow \infty \ea \right. \\
t &=& \half k \bar\rho . \nonumber
\ee
The first term of eq.(\ref{A_det_1}) has the form of the 't Hooft
interaction \cite{tH76}. The form eq.(\ref{A_det_2}) is equal to
eq.(\ref{A_det_1}) to leading order in $m$.
\par
To determine the effective fermion action, we have to find the saddle 
point of the fermion integral, eq.(\ref{S_average}). We introduce an 
integral representation for the powers of the one--instanton averages,
\be
\lefteqn{\overline{\rm Det}_{N_\pm} 
\; = \; \int \frac{d\lambda_+}{2\pi}
\int \frac{d\lambda_-}{2\pi} \exp \left( 
N_+ \log\left[\frac{N_+}{\lambda_+ V}
\left(\frac{4 \pi^2 \bar\rho^2}{N_c M_1}\right)^{N_f}\right] - N_+ 
\;\; + \;\; (+ \rightarrow -)
\right) } && 
\label{Z_lambda} \\
&\times&  \!\! \intpsi \exp \int \! d^4 x \left( \sum_f^{N_f}
\psi^\dagger_f i \partialslash \psi_f 
\;\; + \;\; \lambda_+ \det \left[ i J_+ (x) 
+ \frac{m N_c}{4\pi^2\bar\rho^2}\right]
\; + \; (+ \rightarrow -) \right) .
\nonumber 
\ee
We now determine the common saddle point of eq.(\ref{Z_lambda}) in the 
variable $\lambda_\pm$ and the fermion field, in the thermodynamic limit, 
$N_\pm \rightarrow \infty , V \rightarrow \infty , N_\pm / V\;{\rm fixed}$.
\par
{\em The case $N_f = 1$}. Let us first consider the case of one quark
flavor, $N_f = 1$. In this case, the exponent in the fermion integral of
eq.(\ref{Z_lambda}) is bilinear in the fermion field,
\be
\int d^4 z \left( i J_\pm (z) + \frac{m N_c}{4\pi^2\bar\rho^2} \right) 
&=& i \fourint{k} \psi^\dagger (k)
\half (1 \pm \gamma_5 ) \psi (k) F^2 (k) \;\; + \;\; 
\frac{m V N_c}{4\pi^2\bar\rho^2} ,
\label{Y_momentum}
\ee
and the fermion integral can be carried out exactly, 
\be
\overline{\rm Det}_{N_\pm} &=& \int \frac{d\lambda_+}{2\pi}
\int \frac{d\lambda_-}{2\pi} \exp\left( - W_1 (\lambda_+ , \lambda_- ) 
\right), 
\nonumber \\
W_1 (\lambda_+ , \lambda_- ) &=& 
- \left( N_+ \log\left[\frac{N_+}{\lambda_+ V}
\left(\frac{4 \pi^2 \bar\rho^2}{N_c M_1}\right) \right] - N_+ 
\; + \; (+\rightarrow -)\right)
\;\; - \;\; 
\frac{m V N_c}{4\pi^2\bar\rho^2} (\lambda_+ + \lambda_- ) \nonumber \\
&-& \Tr\log \left[ \frac{\kslash - \half i \left[ \lambda_+ (1 + \gamma_5 )
+ \lambda_- (1 - \gamma_5 ) \right] F^2 (k)}{\kslash - i m} \right] .
\label{W_lambda}
\ee
(Here, ${\rm Tr}$ includes the trace over momentum space, and color and
spinor indices.) Since the fluctuations of
$\Delta = N_+ - N_-$ are small relative to $N = N_+ + N_-$ in the
thermodynamic limit, it is sufficient to evaluate the effective action
to leading order in $\Delta$. We set
\be
\lambda_\pm &=& M (1 \pm \delta ) ,
\ee
at the saddle point, with $\delta = O(\Delta )$. The values of $M$
and $\delta$ are determined by minimizing the effective potential,
eq.(\ref{W_lambda}). This leads to the condition
\be
4 N_c V\fourint{k} \frac{M^2 F^4 (k)}{M^2 F^4 (k) + k^2}
&=&  N - \frac{m M V N_c}{2 \pi^2\bar\rho^2} .
\label{self}
\ee
This equation is identical to the self--consistency condition 
of \cite{DP86}. Note that $M$ tends to a finite value in the 
thermodynamic limit. Parametrically,
\be
M &\propto& \left(\frac{N}{V N_c} \right)^{1/2} \bar\rho .
\label{M_parametric}
\ee
Furthermore,
\be
\delta &=& \frac{2\pi^2\bar\rho^2 \Delta}{m M V N_c} ,
\label{delta_sp}
\ee
at the saddle point. The effective fermion action, eq.(\ref{Z_eff}), is 
thus determined as
\be
S_{\rm eff} [\psi^\dagger , \psi ]_{N_f = 1} &=& \fourint{k} 
\psi^\dagger (k)
\left[ \kslash - i M F^2 (k) ( 1 +  \delta \gamma_5 ) \right] \psi (k) .
\label{S_eff_1f}
\ee
It describes fermions with a dynamical momentum--dependent mass,
which is proportional to the square of the wave function of the instanton 
zero mode. For $\Delta \neq 0$, {\em i.e.}, for an unequal number of 
$I$'s and $\bar I$'s, the fermion action shows a parity--violating mass 
term. This term diverges in the chiral limit; we have retained only the 
infrared--singular part and dropped all finite terms. This infrared 
singularity is canceled when averaging over $\Delta$ in the grand 
canonical ensemble and does not affect infrared--stable physical 
quantities (see section 4.2).
\par
{\em The case $N_f > 1$}. In the case of more than one quark flavor,
$N_f > 1$, the instanton--fermion vertex, eq.(\ref{A_det_2}), describes a
many--fermionic interaction ($2 N_f$ fields). In order to perform the 
integral over the fermion field, we have to introduce auxiliary boson 
fields to linearize the exponent of eq.(\ref{Z_lambda}). This can
be done in the large--$N_c$ limit, using the formula
\be
\exp (\lambda \det [i J] ) &=&
\int d\cM \exp\left[ - (N_f - 1) \lambda^{-\frac{1}{N_f - 1}}
(\det\cM )^{\frac{1}{N_f - 1}} + i \tr (\cM J) \right] ,
\label{Det_bosonization}
\ee
which holds in saddle point approximation. Here, $\cM$ is a hermitean 
$N_f \times N_f$--matrix variable. With the help of 
eq.(\ref{Det_bosonization}), eq.(\ref{Z_lambda}) is represented as
\be
\lefteqn{\overline{\rm Det}_{N_\pm}
\; = \; \int \frac{d\lambda_+}{2\pi}
\int \frac{d\lambda_-}{2\pi} \exp \left( 
N_+ \log\left[\frac{N_+}{\lambda_+ V}
\left(\frac{4 \pi^2 \bar\rho^2}{N_c M_1}\right)^{N_f}\right] - N_+ 
\;\; + \;\; (+ \rightarrow -) \right) }
\label{Z_bosonized} \\
&\times& \int \cD\cM_+ \cD\cM_-
\exp \int d^4 x
\left( -(N_f - 1) \lambda_+^{-\frac{1}{N_f - 1}}
(\det\cM_+ (x))^{\frac{1}{N_f - 1}} 
\;\; - \;\; (+ \rightarrow -) \right. \nonumber \\
&& \hspace{5cm} \left. \;\; + \;\; \frac{N_c}{4\pi^2\bar\rho^2} 
\tr \left[ m (\cM_+ (x) + \cM_- (x)) \right] \right) \nonumber \\
&\times& \intpsi \exp \int d^4 x \left( \sum_f^{N_f} \psi^\dagger_f
i \partialslash \psi_f \;\; + \;\; i \tr \left[
\cM_+ (x) J_+ (x) + \cM_- (x) J_- (x) \right] \right) .
\nonumber
\ee
One can now integrate over the fermion field. Furthermore, one may
integrate over the auxiliary variables, $\lambda_\pm$, in saddle point
approximation, and represent the partition function as
\be
\overline{\rm Det}_{N_\pm} &=& \int \cD\cM_+ \cD\cM_-
\exp \left( - W[\cM_+ , \cM_- ] \right) .
\label{Z_W_eff}
\ee
For constant meson fields, $\cM_\pm (x) \equiv \cM_\pm$, the saddle point 
condition for $\lambda_\pm$ is
\be
\lambda_\pm &=& \left(\frac{V}{N_\pm}\right)^{N_f - 1} \det \cM_\pm .
\label{lambda_M}
\ee
Assuming the vacuum meson field to be diagonal in flavor,
$\cM_\pm = {\rm diag} (\cM_{\pm 1}, \ldots \cM_{\pm N_f})$,
the effective potential is obtained as
\be
W(\cM_+ , \cM_- ) &=& - \sum_f^{N_f} \left( N_+ 
\log\left[\frac{N_+}{\cM_{+f} V}
\left(\frac{4 \pi^2 \bar\rho^2}{N_c M_1}\right) \right]
- N_+  \;\; + \;\; (+ \rightarrow -) \right)
\label{W_eff}\\
&-& \frac{N_c V}{4\pi^2\bar\rho^2} 
\sum_f^{N_f} m_f (\cM_{+ f} + \cM_{- f}) \label{W_Nf} \nonumber \\
&-& \sum_f^{N_f}
\Tr\log \left[ \frac{\kslash - \half i \left[ \cM_{+ f} (1 + \gamma_5 )
+ \cM_{- f} (1 - \gamma_5 ) \right] F^2 (k)}{\kslash - i m}
\right] .
\nonumber
\ee
To leading order in $N_c$, the fermion determinant is simply the
product of the determinants for the different fermion flavors, and
the effective potential is the sum of the potentials for one flavor,
eq.(\ref{W_lambda}). Setting
\be
\cM_{\pm f} &=& M_f (1 \pm \delta_f ) ,
\label{M_gen}
\ee
the minimization of eq.(\ref{W_Nf}) thus leads to the same conditions
as for $N_f = 1$, eqs.(\ref{self}, \ref{delta_sp}), for each flavor.
Substituting this vacuum value of the meson field in eq.(\ref{lambda_M}), 
and returning to the representation eq.(\ref{Z_lambda}), we obtain the 
effective fermion action in the chiral limit,
\be
S_{\rm eff} [\psi^\dagger , \psi ] &=& - \left( \int d^4 x
\sum_f^{N_f} \psi^\dagger_f i\partialslash \psi_f
\; + \; (1 + \delta ) Y_+ \; + \; (1 - \delta) Y_- \right) ,
\label{S_eff} \\
Y_\pm &=& \left(\frac{2V}{N}\right)^{N_f - 1} (i M)^{N_f} 
\int d^4 x \, \det J_\pm (x) ,
\label{Y} \\
\delta &=& \sum_f^{N_f} \delta_f
\; = \; \frac{2\pi^2\bar\rho^2 \Delta}{M V N_c} 
\left( \sum_f^{N_f} m_f^{-1}
\right) .
\label{delta_gen}
\ee
This general formula includes as a special case the result for $N_f = 1$,
eq.(\ref{S_eff_1f}). The dynamical fermion mass is the same as
for $N_f = 1$. (Again, we have dropped $O (m)$ terms in the
$CP$--symmetric part of the action, so that $M_1 = \ldots = M_{N_f} = M$.)
\subsection{Fermions and the topological susceptibility}
In deriving the effective fermion action for given $N_\pm$, we have used
the instanton size distribution of pure gluodynamics, eq.(\ref{d_rho}),
an approximation justified by the formal parameter $N_f/N_c$.
Indeed, the modifications of the size distribution due to 
the low--energy fermions are only quantitative. However, when considering 
the grand partition function, eq.(\ref{GPF}), there is one important aspect 
in which fermions qualitatively alter the picture. The presence of 
dynamical light fermions modifies the probability of topological 
fluctuations, $\Delta = N_+ - N_- \neq 0$, in a principal way:
$\langle \Delta^2 \rangle$ must vanish in the chiral limit. This
remarkable property is a consequence of the $U(1)_A$--anomaly of QCD.
In the following, we derive the topological susceptibility from the
QCD anomaly equation and the chiral Ward identities. We then show, that
the instanton vacuum leads to an identical result in the chiral limit.
\par
The anomalous divergence of the isosinglet axial current operator
is given by
\be
\sum_f^{N_f}
\partial^\mu (\bar\psi_f \gamma_\mu \gamma_5 \psi_f ) &=&
\frac{N_f}{16\pi^2} F\Fdual \; + \; 2 i \sum_f^{N_f} m_f \bar\psi_f 
\gamma_5 \psi_f .
\label{anomaly}
\ee
For the time--ordered product of two topological charge densities, one
obtains the Ward identity \cite{WV79}
\be
- 2 i \sum_f^{N_f} m_f \int d^4 x \, T 
\langle 0 | F\Fdual (x) , \,\bar\psi_f \gamma_5 \psi_f (0) | 0 \rangle
&=& \frac{N_f}{16\pi^2} \int d^4 x \, T \langle 0 | F\Fdual (x) ,
F\Fdual (0) | 0 \rangle . 
\nonumber \\
\label{ward}
\ee
To determine the topological susceptibility, we saturate
the left--hand side of eq.(\ref{ward}) with pseudoscalar meson states.
In general, there are $N_f^2$ pseudo--scalar bosons related to the 
breaking of $U(N_f ) \times U(N_f )$, of which $N_f^2 - 1$ become massless 
in the chiral limit, (Goldstone bosons), while one singlet state remains
massive. The couplings of these states to the
non-singlet and singlet axial currents can be written as
\be
q_\mu \, \langle 0 | \,\bar\psi t^a \gamma_\mu \gamma_5 \psi (0)
\, | \pi_A (q) \rangle &\equiv& F_\pi M_A^2\, X^a_A , \hspace{2em}
(a = 0, \ldots , N_f^2 -1 ) .
\label{X}
\ee
Here, the flavor generators, $t^a$, are normalized according to
\be
\tr \left[ t^a t^b \right] &=& \half\delta^{ab}.
\ee
A unitary mixing matrix, $X^a_A$, takes into account that,
for $m\neq 0$, mass and flavor eigenstates are in general not identical. 
(The index $A$ runs over the $N_f^2$ ``nonet'' meson states.)
We now need the matrix elements of the operators on the left--hand
side of eq.(\ref{ward}) between the vacuum and one--meson states. 
The topological charge density, $F\Fdual$, is a flavor singlet, and its 
matrix element can be parametrized as 
\be
\langle 0 | F\Fdual (0) | \pi_A \rangle &\equiv&
\frac{\kappa F_\pi}{\sqrt{2N_f}} X^0_A ,
\label{FFdual_me}
\ee
where $\kappa$ is a constant, to be determined below. The matrix
element of the pseudoscalar density, $\bar\psi t^a \gamma_5 \psi$, 
is obtained from the non-singlet Ward identity,
\be
-i \int d^4 x \, T \langle 0 | \,
\bar\psi \, \{ t^a , m \}_+ \gamma_5 \psi (x) , \,
\bar\psi t^b \gamma_5 \psi (0) \, | 0 \rangle
&=& \langle 0 | \, \bar\psi \, \{ t^a , t^b \}_+ \psi (0) \, | 0 \rangle .
\ee
Inserting the pseudoscalar mass states, $\pi_A$, and using the definition 
of the mixing matrix, eq.(\ref{X}), one finds
\be
\langle 0| \bar\psi t^a \gamma_5 \psi |\pi_A \rangle &=&
i \frac{\langle \bar \psi \psi\rangle}{F_\pi} X^a_A .
\label{density_me}
\ee
Here,
\be
\langle \bar\psi \psi \rangle &\equiv& \frac{1}{N_f}\sum_f^{N_f}
\langle \bar\psi_f \psi_f \rangle.
\ee
Finally, the mixing matrix, $X^a_A$, is determined by taking the matrix 
element of the anomaly equation, eq.(\ref{anomaly}), and the 
non-anomalous divergence of the non-singlet axial current,
\be
\partial^\mu ( {\bar\psi} \gamma_\mu \gamma_5 t^a \psi )
&=& i  {\bar\psi} \{ t^a, m \}_+ \gamma_5 \psi
\hspace{2em} (a = 1,\ldots, N_f^2 - 1) ,
\ee
which leads to an equation
\be 
\sum_A M_A^2 \, X^a_A \, X^{b\ast}_A &=& - \frac{2\langle \bar\psi \psi
\rangle}{F_\pi^2} \tr \left[ \{ m, t^a \}_+ t^b \right] 
\;\; + \;\; \kappa \delta^{a0} \delta^{b0} .
\label{mass_matrix}
\ee
\par
Inserting now singlet and non-singlet states into eq.(\ref{ward}),
using eqs.(\ref{FFdual_me}, \ref{density_me}), one obtains
\be
\frac{1}{16\pi^2} \int d^4 x \, T \langle 0 | F\Fdual (x) ,
F\Fdual (0) | 0 \rangle
&=&  \frac{\kappa F_\pi^2}{2 N_f} \left( 1 - \kappa
\sum_A \frac{X^0_A \, X^{0\ast}_A}{M_A^2} \right) .
\label{ward_mesons}
\ee
Solving eq.(\ref{mass_matrix}) for $X^a_A$, one obtains 
after a straightforward calculation
\be
\frac{\langle Q_t^2 \rangle}{V} &=& 
\frac{\kappa \langle \bar\psi \psi \rangle}
{{\displaystyle\frac{2N_f}{F_\pi^2}} \langle \bar\psi \psi \rangle 
- \kappa \left( \displaystyle{\sum_f^{N_f}}
m_f^{-1} \right)} .
\label{QQ_unquenched_full}
\ee
In the chiral limit, $m\rightarrow 0$, this becomes
\be
\frac{\langle Q_t^2 \rangle}{V} &=& - \langle \bar\psi \psi \rangle 
\left(\sum_f^{N_f} m_f^{-1}\right)^{-1} .
\label{QQ_unquenched}
\ee
This replaces the gluodynamics formula, eq.(\ref{QQ}), in the presence 
of light fermions. The topological susceptibility is proportional to 
the harmonic average of the quark masses, which means that it vanishes 
if (at least) one of the fermion flavors becomes massless.
\par
It is interesting to consider eq.(\ref{QQ_unquenched_full}) also
in the ``quenched'' limit, {\em i.e.}, to switch off the fermions before
letting the quark masses go to zero. (Formally, this corresponds to
the limit $N_f \rightarrow 0$, with $\sum m_f^{-1} = O (N_f)$ and
$\langle \bar\psi \psi\rangle = O (1)$.) In this case, 
eq.(\ref{QQ_unquenched_full}) reduces to
\be
\kappa &=& \frac{2 N_f \langle Q_t^2 \rangle_{\rm quenched}}{F_\pi^2 V} .
\label{WV1}
\ee
Taking now the limit $m \rightarrow 0$, the mixing between singlet and 
non--singlet states disappears, and one obtains from 
eq.(\ref{mass_matrix})
\be
\kappa &=& M_{\rm singlet}^2 .
\label{WV2}
\ee
Eqs.(\ref{WV1}, \ref{WV2}) are the well-known Witten--Veneziano formula 
for the topological susceptibility in quenched QCD \cite{WV79}.
\par
In the instanton language, the vanishing of the topological 
susceptibility for $m \rightarrow 0$ can be understood as being due to 
the ``unbalanced'' fermionic zero modes in the fermion determinant at 
$\Delta \neq 0$. To see this explicitly, we calculate the fermion 
determinant in zero mode approximation, eq.(\ref{Z_f_opt}), as a function 
of $\Delta$. Evaluating the integral representation, 
eqs.(\ref{Z_W_eff}, \ref{W_eff}), at the saddle point defined by 
eqs.(\ref{self}, \ref{M_gen}, \ref{delta_gen}), we obtain, to leading 
order in $1/m$,
\be
\overline{\rm Det}_{N_\pm} &\propto&
\exp \left( - \frac{\pi^2 \bar\rho^2 \Delta^2}{N_c M V} 
\sum_f^{N_f} m_f^{-1} \right) .
\ee
On the other hand, the quark condensate, calculated from eq.(\ref{W_eff})
at $\Delta = 0$, is 
\be
\langle \bar\psi \psi \rangle_{\rm Minkowski} &=&
-i \langle \psi^\dagger \psi \rangle_{\rm Euclidean} \;\; = \;\;
\frac{1}{N_f} \sum_f^{N_f}
\left( - \frac{1}{V} \frac{\partial}{\partial m_f} \log 
\overline{\rm Det}_{N_\pm} \right)_{\Delta = 0}
\; = \; - \frac{N_c M}{2\pi^2 \bar\rho^2}. 
\nonumber \\
\ee
One thus sees that the $\Delta$--distribution described by the 
fermion determinant of eq.(\ref{Z_f_opt}) precisely realizes the theorem
eq.(\ref{QQ_unquenched}),
\be
\langle \Delta^2 \rangle &=& \frac{N_c M V}{2\pi^2 \bar\rho^2}
\left( \sum_f^{N_f}  m_f^{-1} \right)^{-1}
\; = \; - V \langle \bar\psi \psi \rangle
\left( \sum_f^{N_f}  m_f^{-1} \right)^{-1} .
\label{Delta_unquenched}
\ee
Note that the $\Delta$--distribution corresponding to the full 
variational partition function, eq.(\ref{Z_f_opt}), including the gluonic 
part,
\be
P(\Delta ) &\propto& Z^{\,\rm fermions}_{N_\pm} \; = \;
Z_{N_\pm} \overline{\rm Det}_{N_\pm} \; \propto \;
\exp \left( - \frac{\Delta^2}{2 N} \right)
\exp \left( \frac{\Delta^2}{2 V \langle \bar\psi \psi \rangle}
\sum_f^{N_f}  m_f^{-1} \right) ,
\label{Z_Delta}
\ee
is dominated by the one from the fermion determinant, since
\be
\frac{N}{V} &\gg& - \langle \bar\psi \psi \rangle
\left(\sum_f^{N_f} m_f^{-1}\right)^{-1} ,
\ee
in the chiral limit.
\par
The variational partition function of independent instantons,
eq.(\ref{Z_f_opt}), including the fermion determinant, thus correctly 
describes the dispersion of topological vacuum fluctuations, which in QCD 
is governed by the $U(1)_A$ anomaly. Contrary to the case of the trace 
anomaly, which is realized already in the quenched approximation (see 
section 2.1), the inclusion of the fermion determinant is essential here. 
In section 4, we shall see that the independent instanton partition 
function with fermions, eq.(\ref{Z_f_opt}), can be used also to 
evaluate matrix elements of the topological charge in hadronic states 
consistently with the $U(1)_A$ anomaly.
\section{Evaluating gluon operators}
\setcounter{equation}{0}
\renewcommand{\theequation}{\arabic{section}.\arabic{equation}}
\subsection{Effective operators for gluons}
To extract information about hadrons, one has to evaluate correlation 
functions of ha\-dro\-nic currents, together with some operators 
consisting of fermion and gauge fields. The prescription for calculating 
vacuum averages of functions of the gauge field and the fermion field is 
defined by the grand canonical instanton partition function with 
fermions, with the fixed--$N_\pm$ partition function given by the 
variational approximation, eq.(\ref{Z_f_opt}). The grand canonical 
averaging is performed in two steps. First, for a fixed number of $I$'s 
and $\bar I$'s, one integrates over the collective coordinates of the 
instantons and over the fermion field, taking into account the 
dynamically generated quark mass. In the final step, one then averages 
over the number of $I$'s and $\bar I$'s according to their dispersion in 
the medium. In this section, we investigate fixed--$N_\pm$ (canonical)
averages of operators, specifically of such involving gluon fields. The
averaging over the number of $I$'s and $\bar I$'s will be discussed in
the following section.
\par
Let us consider first averages of operators involving only
fermion fields. For example, the baryon correlation function is defined
as the vacuum average of two baryon currents, consisting of $N_c$ fermion 
fields coupled to a color singlet,
\be
J_N (x) &=& \frac{1}{N_c !} \, \varepsilon^{\alpha_1 \ldots \alpha_{N_c}} 
\, \Gamma_{s_1 f_1 , \ldots s_{N_c} f_{N_c}} \, 
\psi^{\alpha_1}_{s_1 f_1}(x)
\ldots \psi^{\alpha_{N_c}}_{s_{N_c} f_{N_c}}(x) ,
\label{J_Gamma}
\ee
($\Gamma$ is a spin--flavor matrix) \cite{DPP88}. For a fixed number of 
instantons, $N_\pm$, the normalized average over the instanton 
coordinates and the fermion fields can be represented as an integral with 
the effective fermion action, eq.(\ref{S_eff}),
\be
\lefteqn{
\llangle J_N (x_1 ) J_N^\dagger (x_2 ) \rrangle_{{\rm fixed}-N_\pm}}
&& \nonumber \\
&\equiv& \overline{\rm Det}_{N_\pm}^{-1}
\intpsi \, J_N (x_1 ) J_N^\dagger (x_2 )
\int \prod_I^{N_+ + N_-} \frac{dz_I}{V} dU_I
\exp\left( - \widetilde{S} [\psi^\dagger , \psi ] \right) 
\label{JJ} \\
&=& \frac{{\displaystyle \intpsi J_N (x_1 ) J_N^\dagger (x_2) 
\exp \left( -S_{\rm eff}[\psi^\dagger , \psi ] \right) }}
{{\displaystyle \intpsi
\exp \left( -S_{\rm eff}[\psi^\dagger , \psi ] \right) }}  \;\; \equiv \;\;
\langle J_N (x_1 ) J_N^\dagger (x_2 ) \rangle_{\rm eff} .
\label{JJ_eff}
\ee
Here and in the following, we use the symbol 
$\langle\ldots\rangle_{\rm eff}$
to denote the normalized average of a function of the fermion field
within the effective fermion theory, as defined by eq.(\ref{JJ_eff}).
The fermion integral can be evaluated to the order in $1/N_c$
to which one knows the effective action.
\par
A new situation arises when calculating averages of operators containing
gluon fields. Consider the nucleon matrix element of a gluon operator, 
$\cF [A]$, {\em e.g.}, the local gauge--invariant operators 
$F_{\mu\nu}F_{\mu\nu}$ or $F_{\mu\nu}\Fdual_{\mu\nu}$, or functions 
thereof. The nucleon matrix element is obtained from the correlation 
function of $\cF [A]$ with nucleon currents, eq.(\ref{J_Gamma}),
\be
\lefteqn{\llangle J_N (x_1 ) J_N^\dagger (x_2 ) \cF [A]
\rrangle_{{\rm fixed}-N_\pm}} &&
\label{JJF} \\
&=& \overline{\rm Det}_{N_\pm}^{-1}
\intpsi J_N (x) J_N^\dagger (x_2)
\int \prod_I^{N_+ + N_-} \frac{dz_I}{V} dU_I \, \cF [A(\xi )] \,
\exp \left( - \widetilde{S}[\psi^\dagger , \psi ] \right)  . 
\nonumber
\ee
Here, the function $\cF [A]$ is evaluated with the gauge field of
the $N_\pm$--instanton configuration, eq.(\ref{A_inst}), and
averaged over the instanton coordinates together with the fermion action 
in the instanton background, eq.(\ref{S_zero_mode}). We want to 
represent eq.(\ref{JJF}) as an average within the effective fermion 
theory of the form of eq.(\ref{JJ_eff}):
\be
\lefteqn{\llangle J_N (x_1 ) J_N^\dagger (x_2 ) \cF [A] 
\rrangle_{{\rm fixed}-N_\pm}} && 
\label{F_eff_def} \\
&\equiv& \frac{{\displaystyle \intpsi J_N (x_1 ) J_N^\dagger (x_2 ) \,\,
\effop{\cF}[\psi^\dagger , \psi ] \,\,
\exp ( -S_{\rm eff}[\psi^\dagger , \psi ] )}}
{{\displaystyle \intpsi \exp (- S_{\rm eff}[\psi^\dagger , \psi ] ) }} 
\nonumber \\
&=& \llangle J_N (x_1 ) J_N^\dagger (x_2 ) 
\effop{\cF}[\psi^\dagger , \psi ] \rrangle_{\rm eff} . \nonumber 
\ee
Here, $\effop{\cF}[\psi^\dagger , \psi ]$ is a (generally non-local) 
fermion operator, which represents the gluon operator, $\cF [A]$, in the 
effective fermion theory. We shall call 
$\effop{\cF}[\psi^\dagger , \psi ]$ the effective operator to $\cF [A]$ 
and denote it by the symbol of the gluon operator put in quotation marks. 
As with the effective fermion action, eq.(\ref{S_eff}), the definition 
eq.(\ref{F_eff_def}) is understood in the sense of the $1/N_c$--expansion, 
{\em i.e.}, the effective operator is defined only in the vicinity of the 
saddle point of the effective fermion theory.
\par
There are several advantages in performing the averaging in the order
of eq.(\ref{F_eff_def}), integrating over instanton coordinates first.
One obtains a simple and transparent representation of
correlation functions with gluon operators. Furthermore, once the
effective operator for a given gluon operator has been determined,
it can be inserted in correlation functions of many different hadronic 
currents. The resulting fermionic correlation functions can 
be evaluated in the chiral soliton picture of the nucleon \cite{DPP88}.
\par
To construct the effective operator, we have to perform in eq.(\ref{JJF})
the integral over the coordinates of the $N_\pm$ instantons, and cast the
remaining fermion integral in the form of an integral with the effective
fermion action, eq.(\ref{S_eff}). Since the effective operator is defined
only in combination with the effective fermion action, it is important
that eq.(\ref{JJF}) is evaluated consistently with the integral defining
the effective fermion action, eq.(\ref{Z_eff}), using the same zero--mode 
approximation for the fermion propagator, eq.(\ref{S_zero_mode}).
\par
For an $N_\pm$--instanton configuration, eq.(\ref{A_inst}), the
function of the gauge field, $\cF [A]$, can be decomposed in
one--instanton and multi--instanton contributions,
\be
\cF [A(\xi )] &=& \sum_I^{N_+} \cF [A_+(\xi_I )] \, + \, 
\sum_{\bar I}^{N_-} \cF [A_- (\xi_{\bar I})]
\;\; + \;\; \mbox{\rm 2-inst.} \;\; + \;\; \mbox{\rm 3-inst.} \;\; 
+ \;\; \ldots
\label{F_inst}
\ee
When averaged over configurations of independent instantons, the
multi--instanton terms are proportional to additional powers of the 
packing fraction of the instanton medium, $\bar\rho^{4} N/V$, relative 
to the one--instanton terms. As the instanton medium is dilute, these 
contributions are numerically small. We retain only the one--instanton 
contributions in eq.(\ref{JJF}). It will be seen that this is consistent 
with the approximations made in deriving the effective fermion theory.
\par
In the following, let us consider for simplicity eq.(\ref{JJF}) without 
the baryon currents, {\em i.e.}, the vacuum average of $\cF [A]$. With the 
zero mode approximation for the fermion propagator, eq.(\ref{S_zero_mode}), 
we obtain in analogy to eq.(\ref{S_average})
\be
\lefteqn{ \langle \cF [A] \rangle_{{\rm fixed}-N_\pm} } &&
\nonumber \\
&=& \overline{\rm Det}_{N_\pm}^{-1}
\intpsi \int \prod_I^{N_+ + N_-} \frac{d^4 z_I}{V} dU_I
\left( \sum_I^{N_+ + N_-} \cF [A_\pm (\xi_I )] \right) 
\exp\left( -\widetilde{S} [\psi^\dagger , \psi ] \right) \\
&=& \overline{\rm Det}_{N_\pm}^{-1}
\intpsi \exp \left( \sum_f^{N_f} \int d^4 x\, \psi^\dagger_f 
i\partialslash \psi_f \right) \nonumber \\
&& \times \left( N_+ \, W_{\cF +} \, W_+^{N_+ - 1} \, W_-^{N_-}
\; + \; N_- \, W_{\cF -} \, W_+^{N_+} \, W_-^{N_- - 1} \right) .
\label{F_average} 
\ee
Again we have used the fact that the instantons are independent. Here, 
$W_\pm$ are the free one--instanton averages, eq.(\ref{A}), while 
$W_{\cF\pm}$ denote the one--instanton averages with an insertion of the 
function $\cF [A]$, evaluated with the gauge field of $I (\bar I)$
\be
W_{\cF \pm}[\psi^\dagger , \psi ] &\equiv& 
(-1)^{N_f} \int \frac{d^4 z_{I(\bar I)}}{V} dU_{I(\bar I)}
\; \cF [A_\pm (\xi_{I(\bar I)})] \; \prod_f^{N_f}
V^{I(\bar I)}_\pm [\psi^\dagger_f , \psi_f ] 
\;\; + \;\; O(m) . \;\;
\label{A_F}
\ee
Here, $\cF [A]$ is averaged over the collective coordinates of one 
$I (\bar I)$ together with the zero mode fermion vertex, eq.(\ref{V_I}), 
which leads to a coupling of the gluon operator to the fermion fields. In 
the important case that $\cF [A]$ is a color--singlet function of the 
gauge field, it is independent of the color orientation of the instanton. 
The average over orientations is thus the same as in the free 
instanton--fermion vertex, eq.(\ref{A_det_1}), and leads to a fermion 
vertex similar to the 't Hooft interaction,
\be
W_{\cF \pm} &=& i^{N_f} \left(\frac{4\pi^2 \bar\rho^2}{N_c M}\, 
\frac{N}{2 V} \right)^{N_f} \frac{2}{N} \, Y_{\cF\pm} , \\
Y_{\cF\pm}[\psi^\dagger , \psi ] &=&
\left( \frac{2 V}{N} \right)^{N_f - 1} (i M)^{N_f} \int 
d^4 z_{I(\bar I)} \; \cF [A_\pm (z_{I(\bar I)}, 
U_{I(\bar I)} = 1, \bar\rho )] \; \det J_\pm (z) . \hspace{1em}
\label{Y_F}
\ee
(The prefactors have been chosen in analogy to the definition of the
vertices, $Y_\pm$, eq.(\ref{Y}).) If $\cF [A]$ is not a color singlet, 
which may be the case if it is part of a mixed fermion--gluon operator
the average over color orientations in eq.(\ref{Y_F}) results in fermion 
vertices with non--singlet color structure.
\par
We want to represent the fermion integral of eq.(\ref{F_average})
as an integral with the effective fermion action, eq.(\ref{S_eff}).
The effective action was derived from a product of $N_+$ $I$'s and
$N_-$ $\bar I$'s. In eq.(\ref{F_average}), however, only $N_\pm - 1$ 
of the $I (\bar I)$'s are ``free'', while one $I(\bar I)$ is connected 
to the operator, $\cF [A]$. The effective operator in the sense of
eq.(\ref{F_eff_def}) can thus not be immediately identified with
the vertices $Y_{\cF\pm}$, eq.(\ref{Y_F}). Rather, it is of the form
\be
\effop{\cF}[\psi^\dagger , \psi ] &=& N_+ \, Y_{\cF +} \, \cR_+
\;\; + \;\; N_- \, Y_{\cF -} \, \cR_- ,
\label{F_eff}
\ee
where $\cR_\pm [\psi^\dagger , \psi ]$ are functions of the
fermion field, which compensate the fact that we have ``taken out'' an 
$I (\bar I)$ from the $N_\pm$ $I(\bar I)$'s that make the effective 
action, eq.(\ref{S_average}). 
\par
To determine the factors $\cR_\pm$, we proceed with the evaluation of 
eq.(\ref{F_average}) in analogy to the derivation of the effective action, 
eq.(\ref{S_average}). Using the same integral representation for the
powers of one--instanton vertices, eq.(\ref{Z_lambda}), but this
time for $N_\pm - 1$ ``free'' vertices, it is easy to see that
$\cR_\pm$ must be chosen as $Y_\mp$ times a factor which compensates
the shift in the saddle point compared to the integral defining the 
effective fermion action, eq.(\ref{Z_lambda}). 
After a straightforward calculation one obtains
\be
\cR_\pm [\psi^\dagger , \psi ] &=& \frac{4}{N^2} \, Y_\mp \, \exp
\left[ \left( 1 - \frac{2}{N} (1 + \delta ) Y_+ \right) \; + \;
\left( 1 - \frac{2}{N} (1 - \delta ) Y_- \right) \right] \\
&=& \frac{Y_\mp}{\langle Y_+ \rangle_{\rm eff} 
\langle Y_- \rangle_{\rm eff}} 
\exp \left[ \left( 1 - \frac{Y_+}{\langle Y_+ \rangle_{\rm eff}}\right) 
+ \left( 1 - \frac{Y_-}{\langle Y_- \rangle_{\rm eff}}\right) \right] .
\label{R_pm}
\ee
Here, $Y_\pm [\psi^\dagger , \psi ]$ are the 't Hooft vertices of the 
effective fermion action, eq.(\ref{Y}). In the second equation we have 
used that, to leading order in $1/N_c$, the vacuum average of $Y_\pm$ is
\be
\llangle Y_\pm \rrangle_{\rm eff} &=& \half N ( 1 \mp \delta )
\;\; + \;\; O(\Delta^2 ) ,
\label{Y_vac}
\ee
as follows from a straightforward calculation, using the saddle point 
equation, eq.(\ref{self}). 
\par
The effective operator, as defined by eqs.(\ref{Y_F}, \ref{F_eff}, 
\ref{R_pm}), has an intuitively obvious form. It consists 
of 't Hooft--type vertices describing the interaction of the gluon 
operator with the fermions mediated by one instanton. The interaction 
strength is governed by $M$, the dynamical fermion mass. The 
one--instanton vertices are multiplied by certain universal functions 
of the fermion fields, $\cR_\pm$. The meaning of these functions can be 
elucidated by expanding eq.(\ref{R_pm}) around the saddle point. To 
first order in $Y_\pm - \langle Y_\pm \rangle_{\rm eff}$, one has
\be
\cR_\pm &\simeq& \frac{1}{\langle Y_\pm \rangle_{\rm eff}}
\left( 1 - \frac{Y_\pm - \langle Y_\pm \rangle_{\rm eff}}
{\langle Y_\pm \rangle_{\rm eff}} \right) 
\;\; \simeq \;\; Y_\pm^{-1} ,
\label{symbolic}
\ee
{\em i.e.}, the functions $\cR_\pm$ can be regarded as the ``inverse''
of a 't Hooft vertex in the vicinity of the saddle point. As will be seen 
below, the presence of the functions $\cR_\pm [\psi^\dagger , \psi ]$ in 
eq.(\ref{F_eff}), in addition to normalizing the effective operator, has 
important consequences when considering the connected average of the 
effective operator with other fermionic operators, such as hadronic 
currents.
\par
Let us calculate the average of the effective operator,
eq.(\ref{F2_eff}), in the vacuum of the effective fermion theory.
To leading order in $1/N_c$ only the disconnected
average of the factors of eq.(\ref{F_eff}) needs to be considered.
Furthermore, when averaging the exponential factor of eq.(\ref{R_pm})
we may use
\be
\langle X^n \rangle_{\rm eff} &=& \langle X \rangle_{\rm eff}^n , 
\hspace{2em}
\langle \exp X \rangle_{\rm eff} \; = \; \exp \langle X \rangle_{\rm eff}
\ee
($X$ is an arbitrary function of the fermion fields). With 
eq.(\ref{Y_vac}), we obtain
\be
\langle \cR_\pm \rangle_{\rm eff} &=& 
\langle Y_\pm \rangle_{\rm eff}^{-1}
\;\; = \;\; \frac{2}{N} (1 \pm \delta ) \;\; + \;\; O (\Delta^2 ),
\ee
and thus
\be
\langle \effop{\cF}[\psi^\dagger , \psi ] \rangle_{\rm eff} &=& 
N_+ \frac{\langle Y_{\cF +} \rangle_{\rm eff}}
{\langle Y_+ \rangle_{\rm eff}} \; + \; N_- \frac{\langle Y_{\cF -} 
\rangle_{\rm eff}}{\langle Y_- \rangle_{\rm eff}} .
\label{F_eff_vac}
\ee
As expected, the average of the effective operator is given by the 
normalized average of the one--instanton vertices, multiplied by the 
number of $I$'s and $\bar I$'s in the ensemble.
\par
Let us construct the effective operators corresponding to the simplest 
gauge--invariant gluonic operators, $F_{\mu\nu}^2$ and 
$F_{\mu\nu}\Fdual_{\mu\nu}$. For one $I (\bar I)$,
\be
F_{\mu\nu}^2 (x) &=& 
\frac{192 \rho^4}{\left[ \rho^2 + (x - z)^2 \right]^4},
\hspace{2em}
F_{\mu\nu}\Fdual_{\mu\nu} (x) \;\; = \;\; \pm F_{\mu\nu}^2 (x) ,
\ee
and the corresponding one--instanton vertices, eq.(\ref{Y_F}), are
\be
Y_{F^2, \pm} (x) &=& \left( \frac{2 V}{N} \right)^{N_f - 1}
(i M)^{N_f}
\int d^4 z \, \frac{192 \rho^4}{\left[ \rho^2 + (x - z)^2 \right]^4}
\, \det J_\pm (z) , 
\label{F2_eff} \\
Y_{F\Fdual , \pm} (x) &=& \pm Y_{F^2, \pm} (x) .
\label{FFdual_eff}
\ee
In practice, we are interested only in matrix elements of these operators 
at zero momentum transfer (forward scattering), and it is sufficient to
consider the integral of the local operators over the four--volume. 
The integrals of eqs.(\ref{F2_eff}, \ref{FFdual_eff}) are identical to 
the 't~Hooft vertices, eq.(\ref{Y}), times the action of one instanton,
\be
\int d^4 x \, Y_{F^2, \pm} (x) &=& 32 \pi^2 \, Y_\pm ,
\hspace{2em} \int d^4 x \, Y_{F\Fdual , \pm} (x) \;\; = \;\; 
\pm 32 \pi^2 \, Y_\pm .
\ee
The vacuum averages of the effective operators, eq.(\ref{F_eff_vac}), are
thus
\be
\llangle \int d^4 x \, \effop{F_{\mu\nu}^2} \rrangle_{\rm eff} &=&
32 \pi^2 (N_+ + N_- ) \;\; = \;\; 
\llangle \int d^4 x \, F_{\mu\nu}^2 \rrangle_{{\rm fixed}-N_\pm} ,
\label{F2_eff_average}
\ee
and similarly for $F_{\mu\nu}\Fdual_{\mu\nu}$, with $N_+ + N_-$ replaced 
by $N_+ - N_-$. The average of the effective operator in the vacuum of 
the effective fermion theory is thus identical to the average of the 
original operator in the fixed--$N_\pm$ instanton ensemble, in accordance 
with the definition, eq.(\ref{F_eff_def}).
\par
Consider now vacuum averages of the effective operators with 
hadronic currents. An important property of the effective operator 
$\int d^4 x \, \effop{F_{\mu\nu}^2}$ is that its average together with an 
arbitrary fermionic operator, $\cO [\psi^\dagger , \psi ]$, reduces to a 
disconnected average,
\be
\llangle \cO \int d^4 x \, \effop{F_{\mu\nu}^2} \rrangle_{\rm eff}
&=& \llangle \cO \rrangle_{\rm eff}
\llangle \int d^4 x \effop{F_{\mu\nu}^2} \rrangle_{\rm eff}
\label{F2_O_average} \\
&=& 32 \pi^2 (N_+ + N_- ) \llangle \cO \rrangle_{\rm eff} \left[ 1 +
O \left( \frac{1}{N_c} \right) \right] .
\nonumber
\ee
The connected part of the average, which would be of the same order in 
$1/N_c$ as the $1/N_c$--correction to the disconnected part in 
eq.(\ref{F2_O_average}), is zero. (Here, the terms ``connected'' and 
``disconnected'' refer to the effective fermion theory.) To see this, 
let us calculate explicitly the connected part of eq.(\ref{F2_O_average}). 
The operator $\cO$ can connect to any of the factors in the effective 
operator, eq.(\ref{F2_eff}). The contribution of $\cO$ connecting to the 
functions $\cR_\pm$, eq.(\ref{R_pm}), is calculated using the rules
\be
\langle \cO , X^n \rangle_{\rm eff,\, conn} &=&
n \langle \cO , X \rangle_{\rm eff,\, conn} \,
\langle X \rangle^{n - 1}_{\rm eff} , \\
\langle \cO , \exp X \rangle_{\rm eff,\, conn} &=&
\langle \cO , X \rangle_{\rm eff,\, conn} \left( \frac{d}{dX} \exp X
\right)_{X = \langle X \rangle_{\rm eff}} \; = \;
\langle \cO , X \rangle_{\rm eff,\, conn} \, 
\exp \langle X \rangle_{\rm eff} ,
\nonumber
\ee
which apply to the leading order in $1/N_c$. One easily obtains
\be
\llangle \cO , Y_\pm \cR_\pm \rrangle_{\rm eff,\, conn} &=&
\llangle \cO , Y_\pm \rrangle_{\rm eff,\, conn} 
\llangle \cR_\pm \rrangle_{\rm eff}
+ \llangle Y_\pm \rrangle_{\rm eff} 
\llangle \cO , R_\pm \rrangle_{\rm eff,\, conn} 
\;\; = \;\; 0 ,
\ee
in accordance with the interpretation of $\cR_\pm$ as the ``inverse''
of $Y_\pm$, eq.(\ref{symbolic}). From this it follows that
\be
\langle \cO \int d^4 x \, 
\effop{F_{\mu\nu}^2} \rangle_{\rm eff,\, conn} &=& 0.
\ee
An analogous property holds for connected averages with the operator
$\int d^4 x \, \effop{F_{\mu\nu}\Fdual_{\mu\nu}}$. 
\par
The role of the effective operators $\int d^4 x \, \effop{F_{\mu\nu}^2}$ 
and $\int d^4 x \, \effop{F_{\mu\nu}\Fdual_{\mu\nu}}$, when averaged 
together with other fermionic operators, is simply to multiply the 
average of these operators by $32 \pi^2 (N_+ \pm N_- )$. There
are no other dynamical effects from the insertion of an operator
$\int d^4 x \, \effop{F_{\mu\nu}^2}$ or
$\int d^4 x \, \effop{F_{\mu\nu}\Fdual_{\mu\nu}}$ in a correlation
function in the effective fermion theory. This is what one should expect: 
the original operators $\int d^4 x \, F_{\mu\nu}^2$ and
$\int d^4 x \, F_{\mu\nu}\Fdual_{\mu\nu}$ measure just the number of 
$I$'s and $\bar I$'s in the ensemble, and, by definition, the average of 
the effective operators in the effective theory is equal to the average 
of the original operator in the instanton ensemble with fermions.
\par
The absence of a connected average in eq.(\ref{F2_O_average}) has 
important consequences when passing from canonical to grand canonical 
averages. This property is crucial for realizing the trace and 
$U(1)_A$--anomaly in the matrix elements of the operators 
$\int d^4 x \, F_{\mu\nu}^2$ and $\int d^4 x \, F_{\mu\nu}\Fdual_{\mu\nu}$ 
in hadronic states, as will be seen in section 5.
\subsection{Fluctuations of the numbers of instantons}
The effective fermion action, together with the effective operators
for gluons introduced in the previous section, allows to
calculate normalized averages of operators in the ensemble with
a fixed number of instantons. Physical quantities, however, are given
by averages over the grand canonical averages, described by the grand 
partition function with fermions, eq.(\ref{GPF}). 
To pass from canonical to grand canonical
averages, we have to sum the fixed--$N_\pm$ normalized average,
eqs.(\ref{JJ}, \ref{JJF}), over the distribution of the numbers of $I$'s
and $\bar I$'s in the ensemble. The weight for a configuration with given
$N_+ , N_-$ is given by the value of the fixed--$N_\pm$ partition 
function with fermions. With the variational approximation, 
eq.(\ref{Z_f_opt}),
\be
P(N_+ , N_- ) &\propto& Z^{\,\rm fermions}_{N_\pm} \;\; = \;\; Z_{N_\pm}
\overline{\rm Det}_{N_\pm} .
\ee
(For convenience, we shall assume the distribution $P(N_+ , N_- )$ to be
normalized.) Since we consider only quadratic (gaussian) fluctuations
around the equilibrium values, $N_+  = N_- = \half\langle N \rangle$,
we may instead of $N_\pm$ equivalently average over independent 
distributions of $N = N_+ + N_-$ and  $\Delta = N_+ - N_-$.
\par
The normalized grand canonical average of a fermionic operator, 
$\cO [\psi^\dagger , \psi ]$, is defined as
\be
\llangle \cO  [\psi^\dagger , \psi ] \rrangle &\equiv& \sum_{N_+ , N_-}
P(N_+ , N_- ) \, \llangle \cO  [\psi^\dagger , \psi ] \rrangle_{\rm eff} .
\label{grand_fermion}
\ee
For a gluon operator, $\cF [A]$, the only difference is that the 
canonical average is calculated with the help of the corresponding 
effective operator,
\be
\llangle \cF [A] \rrangle &\equiv& \sum_{N_+ , N_-}
P(N_+ , N_- ) \, \llangle \effop{\cF} [\psi^\dagger , \psi ] 
\rrangle_{\rm eff} .
\label{grand_gluon}
\ee
Here, it is assumed that the fixed--$N_\pm$ averages over the effective 
fermion theory have been evaluated as functions of $N_\pm$, in the 
vicinity of $N_+  = N_- = \half\langle N \rangle$ and to first order in 
$\Delta = N_+ - N_-$.
\par
In particular, with this definition, the grand canonical average
of $\int d^4 x \, F_{\mu\nu}^2$, using eq.(\ref{F2_eff_average}), is
\be
\llangle \frac{1}{32\pi^2}\int d^4 x \, F_{\mu\nu}^2 \rrangle &=& 
\sum_{N_+ , N_-} P(N_+ , N_- ) \, \llangle \frac{1}{32\pi^2} 
\int d^4 x \, \effop{F_{\mu\nu}^2} [\psi^\dagger , \psi ] 
\rrangle_{\rm eff} \nonumber \\
&=&  \sum_{N_+ , N_-} P(N_+ , N_- ) \, (N_+ + N_-) \; = \;
\langle N \rangle .
\ee
The effective operator technique reproduces the usual value of 
the vacuum gluon condensate. In the same way one obtains that the
grand canonical average of $\int d^4 x \, F_{\mu\nu}\Fdual_{\mu\nu}$ is 
zero.
\par
The relative width of fluctuations of $N - \langle N \rangle$ and
$\Delta$ is inversely proportional to the size of the system,
{\em cf.}\ eqs.(\ref{DN}, \ref{Z_Delta}). Thus, fluctuations have no
effect on averages of intensive quantities in the thermodynamic limit.
For example, the correlation function of two baryon currents in the
grand canonical ensemble simply reduces to the one in the canonical
ensemble, eq.(\ref{JJ}), evaluated at the equilibrium value of $N_\pm$,
\be
\llangle J_N (x_1 ) J_N^\dagger (x_2 ) \rrangle &=&
\llangle J_N (x_1 ) J_N^\dagger (x_2 ) \rrangle_{\rm eff}\;
\rule[-.3cm]{.15mm}{.7cm} \; 
\rule[-.2cm]{0mm}{0cm}_{N = \langle N \rangle , \; \Delta = 0}
\;\; + \;\; O \left( \frac{1}{\langle N \rangle}, \frac{1}{V} \right) .
\ee
As a consequence, the infrared--singular term of the effective fermion
action proportional to $\Delta$, eq.(\ref{S_eff}), does not affect
free hadron correlation correlation functions, as it should be.
However, fluctuations of $N$ and $\Delta$ lead to non-trivial 
consequences, when considering connected grand canonical averages of 
fermionic currents with extensive operators, such as 
$\int d^4 x \, F_{\mu\nu}^2$ and 
$\int d^4 x \, F_{\mu\nu}\Fdual_{\mu\nu}$.
We shall now discuss these effects, considering separately 
non-topological ($N - \langle N \rangle$) and topological ($\Delta$) 
fluctuations.
\par
{\em Non-topological fluctuations}. Let us consider the connected 
grand canonical average of an arbitrary fermionic operator, 
$\cO [\psi^\dagger , \psi ]$, with the operator 
$\int d^4 x\, F_{\mu\nu}^2$. According to 
eqs.(\ref{grand_fermion}, \ref{grand_gluon}),
\be
\lefteqn{ \llangle \cO \,
\frac{1}{32\pi^2} \int d^4 x \, F_{\mu\nu}^2 \rrangle 
- \llangle \cO \rrangle \llangle \frac{1}{32\pi^2} \int d^4 x \, 
F_{\mu\nu}^2 \rrangle } && \nonumber \\
&=& \sum_N P(N) \llangle \cO \, \frac{1}{32\pi^2} \int d^4 x \, 
\effop{F_{\mu\nu}^2} \rrangle_{\rm eff} 
\nonumber \\
&& - \left( \sum_N P(N) \llangle \cO \rrangle_{\rm eff} \right)
\left( \sum_N P(N) \llangle \frac{1}{32\pi^2} \int d^4 x \, 
\effop{F_{\mu\nu}^2} \rrangle_{\rm eff} \right)
\nonumber \\
&=& \sum_N P(N) (N - \langle N \rangle ) \llangle \cO \rrangle_{\rm eff}
\;\;\; = \;\;\;
\frac{\langle N^2 \rangle - \langle N \rangle^2}{\langle N \rangle}
\left( N \frac{d}{dN} \langle\cO\rangle_{\rm eff}
\right)_{N = \langle N \rangle} .
\label{F2_O_grand}
\ee
(It is understood that $\Delta = 0$; we do not write the
averaging over $\Delta$ here.) Here, we have employed the property 
eq.(\ref{F2_O_average}) for the fixed--$N$ average. In the last equation, 
we have used the fact that the relative width of the fluctuations,
$\langle (N - \langle N \rangle )^2 \rangle / \langle N \rangle$, becomes
small in the thermodynamic limit. (In addition, the width
is proportional to $1/N_c$, eq.(\ref{DN})). In eq.(\ref{F2_O_grand}) it 
is implied that $\langle\cO\rangle_{\rm eff}$ is first evaluated by 
integrating over the fermion fields for fixed $N$, and then differentiated 
with respect to $N$. This differentiation also includes the dependence of 
$\langle\cO\rangle_{\rm eff}$ on $N$ through the instanton size, 
$\bar\rho$, which is related to $N/V$ according to eq.(\ref{selfcons}). 
Since the only scale in the effective fermion theory is the instanton 
density, $N/V$, the operator $N (d/dN)$ measures the total scale dependence 
of $\langle\cO\rangle_{\rm eff}$. In this sense, it plays the role of
$\Lambda (d/d\Lambda)$ in QCD.
\par
Taking the $N$--distribution from gluodynamics (quenched approximation),
eq.(\ref{DN}), eq.(\ref{F2_O_grand}) becomes
\be
\llangle \cO \, \frac{1}{32\pi^2} \int d^4 x \, F_{\mu\nu}^2 \rrangle 
- \llangle \cO \rrangle
\llangle \frac{1}{32\pi^2} \int d^4 x F_{\mu\nu}^2 \rrangle
&=& \frac{4}{b} \left( N \frac{d}{dN}
\langle\cO\rangle_{\rm eff} \right)_{N = \langle N \rangle} .
\label{N_diff}
\ee
We have reduced the connected grand canonical correlation
function to the $N$--derivative of a fixed--$N$ correlation function,
calculable in the effective fermion theory at equilibrium, 
$N = \langle N \rangle$.
Eq.(\ref{N_diff}) is the statement of the fact that the renormalization
(scale dependence) properties of QCD are realized in our approach.
As a special case, eq.(\ref{N_diff}) contains the low--energy theorem
of scale invariance for the vacuum averages of local fermionic
operators \cite{N80,DP84_1}. If $\cO$ is a local operator,
then, as $N/V$ is the only scale in the effective fermion theory,
\be
\llangle \cO \rrangle_{\rm eff} &\propto& 
\left(\frac{N}{V}\right)^{d/4} ,
\hspace{2em}
\left( N \frac{d}{dN} \langle\cO\rangle_{\rm eff} 
\right)_{N = \langle N \rangle}
\; = \; \frac{d}{4} \llangle \cO \rrangle_{\rm eff} .
\ee
Here, $d$ is the naive dimension of the operator.
\par
We emphasize that eq.(\ref{N_diff}) is valid also for non--local 
fermionic operators, in particular, for operators of the form
$\cO = J_N (x_1 ) J_N^\dagger (x_2 )$. Such operator have an intrinsic 
scale, defined by the distance of the two points, $x_1 , x_2$, and the 
vacuum average is no longer a homogeneous function of
$N/V$. However, eq.(\ref{N_diff}) still leads to useful results
in the limit of large distance, $|x_1 - x_2| \rightarrow \infty$. In 
this regime, the correlation function possesses an asymptotic expansion,
the coefficients of which are dimensionful and therefore proportional
to powers of the instanton density. This fact can be used
to derive the nucleon matrix element of $\int d^4 x \, F_{\mu\nu}^2$ 
(see section 5).
\par
To summarize, a correct realization of the scale dependence properties
of QCD is achieved in our approach already in quenched approximation,
neglecting the influence of the fermion determinant on the distribution 
of $N$. This is different for topological fluctuations, which we shall 
now discuss.
\par
{\em Topological fluctuations.}
We now consider correlation functions with the topological
charge operator, $\int d^4 x \, F_{\mu\nu}\Fdual_{\mu\nu}$. In this 
case, the average over fluctuations of $\Delta = N_+ - N_-$ leads to 
non-trivial consequences. In analogy to eq.(\ref{N_diff}), the grand 
canonical average of an arbitrary fermionic operator, 
$\cO [\psi^\dagger , \psi ]$, with the topological charge becomes
\be
\llangle \cO  \frac{1}{32\pi^2} \int d^4 x \, F_{\mu\nu} \Fdual_{\mu\nu} 
\rrangle &=& \sum_\Delta P(\Delta ) 
\llangle \cO \frac{1}{32\pi^2} \int d^4 x \, 
\effop{F_{\mu\nu}\Fdual_{\mu\nu}} \rrangle_{\rm eff}
\nonumber \\
&=& \sum_\Delta P(\Delta ) \, \Delta \llangle \cO \rrangle_{\rm eff}
\;\;\; = \;\;\; \langle \Delta^2 \rangle \left( \frac{d}{d \Delta}
\llangle\cO\rrangle_{\rm eff} \right)_{\Delta = 0} .
\label{Delta_N_diff}
\ee
Here, we have used that $\langle \Delta^2 \rangle^{1/2}/ N$ is small
for large $N$. The derivative with respect to $\Delta$ of the
fixed--$N_\pm$ average in the effective theory can be calculated as
\be
\left(\frac{d}{d\Delta} \llangle\cO\rrangle_{\rm eff} 
\right)_{\Delta = 0}
&=& \llangle \cO \, \frac{dS_{\rm eff}}{d\Delta} \rrangle_{\rm eff} 
\rule[-.3cm]{.15mm}{.7cm} \; \rule[-.2cm]{0mm}{0cm}_{\Delta = 0} , \\
\frac{dS_{\rm eff}}{d\Delta}[\psi^\dagger , \psi ]
&=& \frac{2\pi^2\bar\rho^2}{N_c V M}
\left( \sum_f^{N_f} m_f^{-1} \right) \left( Y_+ - Y_- \right) .
\label{dS_dDelta}
\ee
Here, $d S_{\rm eff}/d\Delta [\psi^\dagger , \psi ]$ is treated as an 
operator insertion, and the average over the effective fermion is now to 
be taken with $\Delta = 0$.
\par
For the dispersion of $\Delta$ in eq.(\ref{Delta_N_diff}), we
must now use the distribution which includes the effect of the fermion
determinant, eq.(\ref{Delta_unquenched}). With this dispersion, one
obtains
\be
\llangle \cO \, \frac{1}{32\pi^2} \int d^4 x \, F_{\mu\nu}\Fdual_{\mu\nu} 
\rrangle
&=& \llangle \cO \left( Y_+ - Y_- \right) 
\rrangle_{\rm eff}|_{\Delta = 0} .
\label{FFdual_Y}
\ee
Notice that the $1/m$--divergence in the chiral limit, which is
present in the effective fermion action for $\Delta \neq 0$, is
compensated by the fact that the dispersion of $\Delta$,
eq.(\ref{Delta_unquenched}), is of order $m$. As a result, 
the right--hand side of eq.(\ref{FFdual_Y}) is finite in the
chiral limit. It would be senseless to average over
$\Delta$ using the quenched distribution,
$\langle\Delta^2 \rangle = \langle N \rangle$ --- the result would
be divergent in the chiral limit. This is an important
difference to the description of non-topological fluctuations.
\par
We now want to demonstrate that eq.(\ref{FFdual_Y}) is consistent with 
the $U(1)_A$ anomaly of QCD. To this end, let us derive the $U(1)$--axial 
current and its divergence in the effective fermion theory
defined by eq.(\ref{S_eff}), at $\Delta = 0$. Consider a 
$U(1)$--chiral rotation with parameter $\varepsilon$,
\be
\psi &\rightarrow& \exp \left( i \varepsilon \gamma_5 \right) \, \psi, 
\hspace{2cm}
\psi^\dagger \; \rightarrow \; \psi^\dagger \exp \left( i
\varepsilon \gamma_5 \right) .
\ee
The change of the effective action, eq.(\ref{S_eff}), gives the
divergence of the axial current,
\be
\int d^4 x\, \partial_\mu J_{5\mu}(x)
&=& \frac{d S_{\rm eff}}{d\varepsilon}
\rule[-.3cm]{.15mm}{.7cm} \; \rule[-.2cm]{0mm}{0cm}_{\varepsilon = 0}
\;\; = \;\; 2 N_f (Y_+ - Y_- ) .
\label{dJ5_eff}
\ee
The current itself is obtained from the fermion kinetic term, as 
usual\footnote{We neglect the contributions to the current due to the 
momentum dependence of the dynamical fermion mass. They are
parametrically small, of order $M\bar\rho$.},
\be
J_{5\mu}(x) &=& \sum_f^{N_f} \psi^\dagger_f (x) \gamma_\mu \gamma_5 
\psi_f (x) .
\ee
Thus, eq.(\ref{FFdual_Y}) can equivalently be written as
\be
\llangle \cO \, \frac{1}{32\pi^2} \int d^4 x F_{\mu\nu}\Fdual_{\mu\nu} 
\rrangle 
&=& \llangle \cO \, \frac{1}{2 N_f}\int d^4 x\, \partial_\mu J_{5\mu} (x)
\rrangle_{\rm eff}
\rule[-.3cm]{.15mm}{.7cm} \; \rule[-.2cm]{0mm}{0cm}_{\Delta = 0} .
\label{FFdual_axial}
\ee
This is precisely the relation which in QCD would follow from
the anomaly equation, eq.(\ref{anomaly}), in the chiral limit.
We have thus shown that our prescription, based on the effective 
fermion theory and the effective operator for 
$\int d^4 x \, F_{\mu\nu}\Fdual_{\mu\nu}$, fully realizes the $U(1)_A$
anomaly at the level of correlation functions.
\par
One should note the different character of the $U(1)_A$ symmetry 
breaking in QCD and the effective fermion theory. In QCD, the symmetry
is broken anomalously, {\em i.e.}, by the high--momentum components
of the fermion field. The effective fermion theory describes only the
low--momentum components of the fermion field, up to a cutoff, 
$\bar\rho^{-1}$, and so, by definition, it has no anomaly. (We have
assumed this in deriving the divergence of the axial current, 
eq.(\ref{dJ5_eff}).) However, in the effective theory the $U(1)_A$
symmetry is broken explicitly, which results in a non-zero
divergence of the axial current. 
\section{Hadronic matrix elements of $F_{\mu\nu}^2$ and 
$F_{\mu\nu}\Fdual_{\mu\nu}$}
\setcounter{equation}{0}
\renewcommand{\theequation}{\arabic{section}.\arabic{equation}}
We now apply the techniques developed in the previous sections to the 
study of the nucleon matrix elements of the operators $F_{\mu\nu}^2$ and 
$F_{\mu\nu}\Fdual_{\mu\nu}$. On general grounds, the nucleon matrix 
elements of these operators can be parametrized as
\be
\langle p'| F_{\mu\nu}^2 (0) | p \rangle &=& A_S (q^2 ) \, m_N \, 
\bar{u}' u ,
\label{F2_nucleon} \\
\langle p' | F_{\mu\nu}\Fdual_{\mu\nu} (0) | p \rangle &=& 
A_P (q^2 ) \, m_N \, \bar{u}' i \gamma_5 u ,
\label{FFdual_nucleon}
\ee
where $q = p' - p$, and $u, u'$ denote nucleon spinors. In QCD, $A_S(0)$
is uniquely determined by the conformal anomaly \cite{SVZ78},
\be
A_S (0) &=& -\frac{32\pi^2}{b} ,
\label{confan}
\ee
whereas $A_P(0)$ is related to the isosinglet axial coupling constant,
$g_A^{(0)}$, by virtue of the axial anomaly equation, eq.(\ref{anomaly}),
\be
A_P (0) &=& \frac{32\pi^2}{N_f} g_A^{(0)},
\label{axan} \\
\langle p | J_{5\mu} | p \rangle &=& 
g_A^{(0)} \bar u \gamma_\mu \gamma_5 u .
\label{g_A}
\ee
We show now that these relations are realized in our approximate 
description. Thus, the renormalization properties of QCD and the axial 
anomaly are correctly taken into account even at the level of hadronic 
matrix elements.
\par
Consider first the operator $F_{\mu\nu}^2 (x)$. The nucleon matrix element 
in our approach is extracted from the large--time limit of the
connected euclidean correlation function,
\be
\llangle J_N ( T ) J_N^\dagger (- T) F_{\mu\nu}^2 (x) \rrangle -
\llangle J_N ( T ) J_N^\dagger (- T) \rrangle
\llangle F_{\mu\nu}^2 (x) \rrangle .
\ee
(Here, ``connected'' refers to the full grand canonical average.) 
For brevity, $J_N (T) \equiv J_N ({\bf 0},  T)$, {\em etc}. Since we are
interested in the zero momentum transfer matrix element, we integrate the
operator $F_{\mu\nu}^2 (x)$ over the euclidean four--volume. One obtains
\be
A_S (0) &=& \lim_{T\rightarrow\infty} \; \frac{1}{2 m_N T} \;
\frac{{\displaystyle \llangle J_N ( T ) J_N^\dagger (- T)
\int d^4 x \, F_{\mu\nu}^2 \rrangle - \llangle J_N ( T ) 
J_N^\dagger (- T) \rrangle \llangle \int d^4 x \, 
F_{\mu\nu}^2 \rrangle }}
{{\displaystyle \llangle J_N ( T ) J_N^\dagger (- T) \rrangle}} .
\nonumber \\
\label{A_S}
\ee
By virtue of the effective version of the ``low--energy theorem'', 
eq.(\ref{N_diff}), this becomes
\be
A_S (0) &=& \lim_{T\rightarrow\infty} \; \frac{1}{2 m_N T} \;
32\pi^2 \; \frac{4}{b} \, \left( N \frac{d}{dN} \log \llangle J_N (T)
J_N^\dagger (-T) \rrangle_{\rm eff} \right)_{N = \langle N \rangle} .
\label{A_S_N_diff}
\ee
For large $T$, the euclidean correlation function decays exponentially,
\be
\llangle J_N (T)
J_N^\dagger (-T) \rrangle_{\rm eff} &\sim&  
\exp \left( -2 T m_N \right) .
\label{exponential}
\ee
Here, $m_N$ is the nucleon mass, as calculated in the effective fermion
theory for a fixed number of instantons, $N$. For dimensional reasons,
\be
m_N &\propto& \left(\frac{N}{V}\right)^{1/4} ,
\label{m_N}
\ee
since the instanton density is the only scale in the effective fermion
theory. (The instanton size, $\bar\rho$, is related to $N/V$ by
eq.(\ref{selfcons}).) With eq.(\ref{exponential}) and eq.(\ref{m_N}), 
we obtain from eq.(\ref{A_S_N_diff})
\be
A_S (0) &=& -\frac{32\pi^2}{b} ,
\label{A_S_trace}
\ee
in agreement with the trace anomaly. This result is again a manifestation
of the fact that the instanton vacuum preserves the renormalization
properties of QCD.
\par
Let us now consider the nucleon matrix element of the topological
charge, eq.(\ref{FFdual_nucleon}). In analogy to eq.(\ref{A_S}),
it is obtained as
\be
A_P (0) &=& \lim_{T\rightarrow\infty} \; \frac{1}{2 m_N T} \;
\frac{{\displaystyle \llangle J_N (T) J_N^\dagger (-T) 
\, F_{\mu\nu}\Fdual_{\mu\nu} (x) \rrangle}}
{{\displaystyle \llangle J_N (T) J_N^\dagger (-T) \rrangle}} .
\ee
With the realization of the axial anomaly in the effective theory, 
eq.(\ref{FFdual_axial}), we have
\be
A_P (0) &=& \lim_{T\rightarrow\infty} \; \frac{1}{2 m_N T} \;
\frac{{\displaystyle \frac{16\pi^2}{N_f}
\llangle J_N (T) J_N^\dagger (-T) \int d^4 x \,
\partial_\mu J_{5\mu} (x) \rrangle_{\rm eff}
\rule[-.3cm]{.15mm}{.7cm} \; \rule[-.2cm]{0mm}{0cm}_{\Delta = 0} }}
{{\displaystyle \llangle J_N (T) J_N^\dagger (-T) 
\rrangle_{\rm eff}
\rule[-.3cm]{.15mm}{.7cm} \; \rule[-.2cm]{0mm}{0cm}_{\Delta = 0} }}
\label{A_P_N_diff} .
\ee
The correlation functions here are to be evaluated in the large--$N_c$
limit, in which the nucleon is described as a chiral soliton in the 
effective fermion theory \cite{DPP88}. In momentum representation, the 
free nucleon correlation function develops a pole at $m_N$, and a nucleon
state emerges in the effective theory. Saturating the correlation 
functions numerator and denominator of eq.(\ref{A_P_N_diff}) with 
one--nucleon states of the effective theory, one finally obtains
\be
A_P (0) &=& \frac{32\pi^2}{N_f} g_A^{(0)} ,
\ee
where $g_A^{(0)}$ now denotes the axial coupling of the nucleon,
as evaluated in the effective fermion theory from the matrix element
of the isosinglet axial current, {\em cf.}\ eq.(\ref{g_A}).
We have thus shown, that the axial anomaly is realized in our
effective description even at the level of hadronic matrix elements:
the nucleon matrix element of $F_{\mu\nu}\Fdual_{\mu\nu}$ reduces
to the nucleon isosinglet axial coupling within our effective 
description.
\par
The nucleon isosinglet axial coupling, $g_A^{(0)}$, has been evaluated
numerically in the chiral quark soliton model of the nucleon, which is
obtained from the above effective fermion theory by 
bosonization \cite{DPP88}.
Calculations give $g_A^{(0)} \sim 0.36$ \cite{BlotzPG93}, which agrees
well with the value obtained in a recent analysis of polarized nucleon 
structure functions\footnote{For a review of the ``proton spin 
problem'', see \cite{Kodaira95}.} \cite{EllisK95}.
\par
It has been suggested by Anselm \cite{Anselm92} that a small value of
$g_A^{(0)}$ could be explained by equating $A_P (0)$ with
$A_S (0)$, and thus $g_A^{(0)}$ with $-N_f/b$, at the 
constituent quark level. He argued that $A_P (0) = A_S (0)$ 
could be obtained from the fact that 
$F_{\mu\nu}\Fdual_{\mu\nu} = \pm F_{\mu\nu}^2$ for one $I (\bar I)$.
Our derivation shows that there is no justification for such a picture.
The fluctuations of the numbers of instantons in the ensemble are crucial 
to consistently describe the matrix elements of $F_{\mu\nu}^2$ and 
$F_{\mu\nu}\Fdual_{\mu\nu}$. For a single constituent quark instead of 
the nucleon, we immediately obtain from eq.(\ref{A_P_N_diff}) 
$g_A^{(0)} = 1$. The fact that a value of $g_A^{(0)}$ considerably 
smaller than $1$ is obtained for the nucleon \cite{BlotzPG93} can be 
understood as an effect of the meson cloud of the nucleon (polarization 
of the Dirac sea by the $N_c$ valence quarks).
\section{Conclusions and outlook}
In this paper we have considered a description of non--perturbative
phenomena in QCD in the form of a grand canonical ensemble of 
instantons. The fluctuations of the number of pseudoparticles play a 
crucial role in realizing the renormalization properties of QCD in the
resulting statistical mechanics system. Fluctuations of $N = N_+ + N_-$
(non-topological) are related to the trace anomaly, while fluctuations 
of $\Delta = N_+ - N_-$ (topological) realize the $U(1)_A$--anomaly. 
The relation between the QCD anomalies and the fluctuations of $N_\pm$ 
has two directions: First, starting from the QCD low--energy theorem of 
scale invariance and the anomalous $U(1)_A$ Ward identities, one can 
predict the dispersion of $N$ and $\Delta$ in the instanton medium from 
first principles Second, as we have seen, the instanton
medium, obtained with a concrete ansatz for the instanton--instanton
and instanton--fermion interaction, exhibits a dispersion of $N$ and
$\Delta$ in accordance with the general principles. An essential 
requirement is a fully self--consistent treatment
of the instanton medium, based on instanton interactions --- any external
stabilizing mechanism would violate the renormalization properties.
\par
Our second aim has been to develop a consistent practical prescription
to evaluate operators --- in particular, gluon operators --- in the
grand canonical ensemble of instantons with fermions. We have invoked
the large--$N_c$ limit, which simplifies the dynamics in two important
aspects: First, the relative width of the fluctuations of the total 
number of pseudoparticles becomes small. Second, in the large--$N_c$ 
limit one can explicitly construct the effective fermion action for 
an ensemble with a fixed number of instantons. This enables us to 
evaluate correlation functions in a two--step procedure: We first 
integrate over instanton coordinates and the fermion field in a 
canonical ensemble, and then average the result over fluctuations of 
$N_\pm$. All fixed--$N_\pm$ correlation functions can be represented
as integrals over the effective fermion theory, in which gluon operators
are replaced by effective fermion operators. We have shown that
in this approach the trace and $U(1)_A$ anomalies are realized at the 
level of hadronic matrix elements of the operators 
$\int F_{\mu\nu}^2$ and $\int F_{\mu\nu}\Fdual_{\mu\nu}$. Consequently, 
the result for the isosinglet axial coupling of the nucleon 
of \cite{BlotzPG93} may be regarded 
as a consistent estimate of the nucleon matrix element of the 
topological charge in the instanton vacuum.
\par
We have established, that our scheme of approximations
(variational treatment of instanton interactions, $N_c$--limit for the
fermions, zero mode approximation) provides a framework for evaluating
gluonic operators, which is consistent with the basic renormalization
properties of QCD. Any attempt to improve these approximations 
({\em e.g.}\ using more sophisticated trial functions, including explicit 
correlations between instantons due to fermions, or going beyond the 
zero--mode approximation for the fermion propagator) should preserve 
these standards of consistency.
\par
An application of the methods developed here is the calculation
of nucleon structure functions, both for leading and non-leading
twist. By the operator product expansion of QCD, the moments of 
the structure functions are related to
forward matrix elements of certain local operators which contain
gluon fields. These matrix elements can be evaluated using the technique 
of effective operators developed here. The moments are obtained at the 
scale set by the instanton background, typically $1/\bar\rho$, and have 
to be put into QCD evolution equations to compare with experimental
information at higher momenta. Work in this direction is in progress.
\\[.2cm]
We are deeply grateful to V.\ Petrov for his help at the preliminary
stage of this work. We also thank P.\ Pobylitsa for discussions,
especially on section 3.2.
\\[.2cm]
This work has been supported in part by the 
Russian Foundation for Fundamental Research, grant 95-07-03662,
by the DFG and COSY (J\"ulich). The work of D.D.\ and M.P.\ is
supported in part by grant INTAS-93-0283. C.W.\ acknowledges the 
hospitality of the Petersburg Nuclear Physics Institute.
%
%

\end{document}